\documentclass[12pt, a4paper]{article}

\usepackage{jheppub}

\usepackage{braket, amsthm, slashed}

\usepackage[utf8]{inputenc}

\usepackage{hyperref} 
\hypersetup{
     colorlinks=true, 
     linkcolor=blue,  
     citecolor=red,  
     filecolor=cyan,
     urlcolor=blue,
     linktoc=section    
          }

\usepackage{xparse}

\numberwithin{equation}{section} 
\usepackage[titletoc]{appendix}

\usepackage{parskip}
\newcommand{\be}{\begin{equation}}
\newcommand{\ee}{\end{equation}}
\newcommand{\beq}{\begin{eqnarray}}
\newcommand{\eeq}{\end{eqnarray}}

\newcommand{\opo}{\mathcal{O}}
\newcommand{\tfd}{\ensuremath{\ket{\text{TFD}}}}
\newcommand{\ketT}{\ket{\text{TFD}}}
\newcommand{\braT}{\bra{\text{TFD}}}

\def\tr{{\rm Tr}}

\def\au{\Theta}
\def\HTFD{H_{\text{TFD}}}

\def\w{{\hat{\omega}}}
\def\Deltae{\sigma_E}

\title{
How to Build the Thermofield Double State
}

\author[a,b,c]{  William Cottrell,}
\author[b, c]{ Ben Freivogel,} 
\author[b]{  Diego M. Hofman,}
\author[b]{ \\ \vspace{4mm}  Sagar F. Lokhande}

\date{}
\affiliation[a]{\vspace{6mm} Physics Department, Stanford University}
\affiliation[b]{Institute for Theoretical Physics, University of Amsterdam}
\affiliation[c]{GRAPPA, University of Amsterdam
\vspace{8mm}}
\emailAdd{wjc72@stanford.edu, benfreivogel@gmail.com, d.m.hofman@uva.nl, sagar.f.lokhande@gmail.com}

\abstract{
Given two copies of any quantum mechanical system, one may want to prepare them in the thermofield double state for the purpose of studying thermal physics or black holes. However, the thermofield double is a unique entangled pure state and may be difficult to prepare. We propose a local interacting Hamiltonian for the combined system whose ground state is approximately the thermofield double. The energy gap for this Hamiltonian is of order the temperature.  Our construction works for any quantum system satisfying the Eigenvalue Thermalization Hypothesis.
}


\begin{document}

\maketitle

\section{Introduction}

Given two copies of any quantum mechanical system, the thermofield double state $\ketT$ is the unique pure state
\begin{equation}
\ket{\text{TFD}} \equiv \frac{1}{\sqrt{Z}} \sum_n e^{-\beta E_n/2} \ket{n}_L \otimes \ket{n}_R \, ,
\end{equation}
where $\ket{n}_{L,R}$ are the energy eigenstates of the individual systems. This is an entangled pure state of the full system with the property that each of the two copies is in the thermal density matrix with temperature $\beta^{-1}$.

In a quantum theory with a gravity dual, this state is dual  to an eternal black hole. Black holes remain poorly understood; it is a matter of debate whether an observer falling into a large black hole falls freely through the horizon, as predicted by the equivalence principle, or encounters a `firewall' at the horizon. 

Despite many papers on this topic, a consensus has not yet been reached.  The primary difficulty is that the notion of a firewall depends on experiences of observers localized near an event horizon.  However, local observables are not believed to exist in quantum gravity.  In principle, all we should discuss is the S-matrix, but, from this data alone it is essentially impossible to decipher the experiences of the brave soul who sailed into the black hole and was long ago scrambled into Hawking radiation. 

A key step forward was taken in \cite{Gao:2016bin, Maldacena:2017axo} where it was realized that by applying a simple perturbation coupling the two sides of an eternal AdS black hole one may make the wormhole traversable. This, in principle, allows us to probe behind the horizon without dealing with issues of bulk locality - all we need to do is send an observer from one side to the other and ask them how they felt.  Susskind has predicted that we will be able to perform experiments of this type `within the next decade or two' \cite{Susskind:2017ney}. The eternal AdS black hole is dual to the thermofield double state of the two boundary CFTs \cite{Maldacena:2001kr, Maldacena:2013xja}. The thermofield double state is also of interest beyond the context of black holes, in the study of thermal field theories. 
 
The first step in performing such experiments is to prepare two copies of a quantum system in the thermofield double (TFD) state. The goal of this article is to propose a simple way to do so. Our approach will be to look for an interacting Hamiltonian for the combined system whose ground state is the  TFD state.  If this `TFD Hamiltonian' can be experimentally realized, and the system has a way to dissipate energy, then the system will eventually approach the TFD state.
 
One might worry that it is difficult to construct the TFD state: it is one state in the very large Hilbert space, and it is not defined in terms of a minimization principle. One simple definition is that the TFD state is generated by evolution in Euclidean time, but we have not been able see how to use this definition in the laboratory. A particular worry is there are many states that look roughly like the TFD state but differ by relative phases,
\begin{equation}
\ket{\text{TFD}_\phi} = \frac{1}{\sqrt{Z}} \sum_n e^{i \phi_n} e^{-\beta E_n/2} \ket{n}_L \otimes \ket{n}_R \, .
\end{equation}
These states have the same thermal density matrix for each of the two subsystems as  the bona fide TFD, but they do not correspond to a bulk dual with a `short' AdS wormhole. Trying to send the teleportee through these bulk geometries will result in disaster. Furthermore, there are of order $\exp(S)$  of these states, while the `real' thermofield double is unique.
  
Our central claim is that in quantum systems satisfying the Eigenvalue Thermalization Hypothesis (ETH), the thermofield double state is in fact the ground state of a relatively simple Hamiltonian.  Schematically, our claim is that a simple Hamiltonian of the form
\begin{equation}
H_{S} \sim H^0_L + H^0_R + \sum_k c_k  \left( \opo_L^k - \opo_R^k\right)^2
\end{equation}
has a ground state that is approximately the thermofield double. Here $H^0_{L, R}$ are the original Hamiltonians of the left and right systems, and the $\opo^k_{L (R)}$ are $any$ operators in the left (right) system. In the Quantum Field Theory context this Hamiltonian is local (in the sense of Effective Field Theory) if the system in question is in the thermodynamic limit. We summarize our results more precisely at the end of this introduction.

Of particular importance in this program is the gap, $\Delta E$, in the Hamiltonian we will be constructing.   This is a measure of how quickly the desired state can be reached and how carefully the experiment must be controlled. 
 It is also indirectly a measure of the complexity of the TFD state since the complexity scales like the time required to reach the state, which scales like $\Delta E^{-2}$ \cite{2002quant.ph..1031F, 2016arXiv161207258P}.  We present evidence that the gap does $not$ become exponentially small in the black hole entropy; in fact, the gap is of order the temperature as long as the number of different operators $\opo^k$ is larger than a few.

\subsection{Summary of Results.}

To be more precise, we define $\ketT$ by
\begin{equation}
\ket{\text{TFD}} \equiv \frac{1}{\sqrt{Z}} \sum_n e^{-\beta E_n/2} \ket{n}_L \otimes \ket{n^*}_R \, .
\end{equation}
Here 
\begin{equation}
\ket{n^*} \equiv \au \ket{n}
\end{equation}
where $\au$ is an anti-unitary operator, such as CPT, that commutes with the original Hamiltonian. This definition is motivated by the path integral construction of the \tfd, which entangles electrons in the right theory with positrons in the left theory, etc.  
Our results can be summarized as follows:
\begin{itemize}

\item  \tfd\ is the ground state of the Hamiltonian 
\begin{align}
\begin{split}
H_{\rm TFD} &= \sum_k c_k d_k^\dagger d_k \, , \\
d_k &\equiv  e^{- \beta (H^0_L + H^0_R)/4} \left(\opo^k_L - \au \opo^{k \dagger}_R \au^{-1} \right) e^{ \beta (H^0_L + H^0_R)/4} \, .
\end{split}
\end{align}
where $H_L^0$ is the original Hamiltonian of the left theory, $\opo^k_{L}$ is $any$ operator in the left system and $\opo^k_R$ is the $same$ operator in the right system.
This Hamiltonian has the exact TFD as the ground state  but it may be quite complicated in the case of interest where the original hamiltonian $H^0$ is strongly coupled. 
\item \tfd\ is the approximate ground state of the simple Hamiltonian
\begin{equation}
\label{eq:simpleh}
H_{S} = \sum_k c_k d_k^\dagger d_k + H_L^0 + H_R^0  \,  , \quad d_k = \opo^k_L - \au \opo^{k \dagger}_R \au^{-1} \, .
\end{equation}
in systems satisfying the Eigenvalue Thermalization Hypothesis (ETH), where the $c_k$ are appropriately chosen positive numbers. 
\item The energy gap between the ground state and the first excited state is of order the temperature scale in systems satisfying ETH, as long as the number different operators $\opo^k$ included in the Hamiltonian is at least a few. Therefore, introducing couplings between a handful of simple operators in the two theories is sufficient to pick out \tfd\ uniquely. 
\end{itemize}

We begin in the next section by giving our general prescription for the TFD Hamiltonian, applying it to a number of examples in section \ref{sec-ex}. In section \ref{EFT} we analyze our Hamiltonian using effective field theory and show that the UV cutoff for the EFT is of order the temperature of the TFD state. In sections \ref{sec-exacth} and \ref{approxH} we analyze in detail the gap for the exact and approximate Hamiltonians. In section \ref{sec-exp} we discuss some sources of error.  In section \ref{qc_connection} we point out a connection between   our setup and certain NP complete problems, as well as offering 
 a speculative interpretation of our construction as a model for a quantum learning algorithm.  More precisely, we are trying to `learn' a state given a small number of operator relations on this state and successful learning may be interpreted as the absence of a firewall.  We close with a number of directions for future research in section \ref{sec-future}.

\subsection*{Previous Work}

During the lengthy interval it took us to complete this work, the interesting paper \cite{Maldacena:2018lmt} by Maldacena and Qi appeared, where the problem of constructing the TFD state was considered and a similar expression for the TFD Hamiltonian was proposed.  In \cite{Maldacena:2018lmt}, the TFD Hamiltonian is similar to equation \eqref{eq:simpleh}, with the interaction term being just $\opo_L \cdot \opo_R$. However, they only study the $q$-body SYK model at large $N$. Further, the coupling constant is taken to be $\opo(1)$, unlike our situation. They show that the ground state of this Hamiltonian is approximately the TFD state, albeit with a small overlap with the real TFD state. This is different from our case, where the overlap is significant. 

A few years ago,  McGreevy and Swingle \cite{2014arXiv1407.8203S, Swingle:2016foj} introduced a general formalism to build mixed states in many-body systems using quantum circuits. They called this formalism \textit{s-sourcery}. Using this formalism, they constructed TFD Hamiltonians for free theories. These Hamiltonians are similar to what we construct here in the free case, and we compare our results where appropriate. Our main goal, however, is to offer a simple proposal for the strongly coupled theories that are of interest for holography.

\section{General Construction of the TFD State}
\label{sec_general_constr}
Our goal is now to provide a prescription for preparing the thermofield double state.  We start with two identical quantum systems and then attempt to construct some interaction such that the ground state of the combined system is precisely $\ket{TFD}$.  As explained in the introduction, we define \tfd\ by
\be
\ketT \equiv \frac{1}{\sqrt{Z}} \sum_{n} e^{-\beta E_{n}/2} \ket{n}_{L} \otimes \ket{{n^*}}_{R} \, ,
\ee
where the $\ket{n}$ are energy eigenstates and $\ket{{n^*}}\equiv \Theta \ket{n}$, with $\Theta$ being an anti-unitary operator such as CPT.

To find this Hamiltonian, let us start with any operator in the left theory  $\opo_L$. Let $\opo_R$ be the corresponding operator in the right theory. Then an operator $d$ of the form,
\begin{equation}
\label{eq:general_d_def}
d \equiv e^{- \beta (H^0_L + H^0_R)/4} \left(\opo_L - \au \opo^{ \dagger}_R \au^{-1} \right) e^{ \beta (H^0_L + H^0_R)/4}
\end{equation}
will annihilate the TFD state,
\begin{equation}
d \ket{\text{TFD}} = 0 \, .
\end{equation}
This is because in the energy eigenbasis, the matrix elements of the two terms in $d$ are equal in magnitude. After some algebra,
\be\label{dtfd}
d \, \ketT =\frac{1}{\sqrt{Z}} \sum_{ij} e^{-\beta (E_i+E_j)/4} \bigg((\opo)_{ij} \ket{i} \ket{j^*} -  ( \opo^{\dagger} )^*_{ji} \ket{i}\ket{j^*}  \bigg) =0\, .
\ee
\noindent with
\be
(\opo)_{ij}  \equiv \bra{i} \mathcal{O} \ket{j}
\ee
The two terms in the parentheses in (\ref{dtfd}) come from the action of $\au$ and are equal. Then the TFD hamiltonian in general will be a sum over such operators,
\begin{equation}
 \label{eq:HTFD_general}
\HTFD = \sum_i  \,  c_i  d_i^\dagger \, d_i \, ,
\end{equation}
where $c_i$ is a set of positive numbers. A useful simplification is that we do not need to include such terms for every operator separately. A state that is annihilated by the $d$ operator built from $\opo_1$ and the $d$ operator built from $\opo_2$ is automatically annihilated by the $d$ operator built from their commutator. 
This is straightforward to show but important for us, so we formalize this as the

\paragraph{Commutator Property}  \hfill

Given $\opo_1$ and $\opo_2$ such that
\be
d_1 \ketT = d_2 \ketT = 0 \, .
\ee
Then $d_3 \ketT=0$, where
\be 
d_3 \equiv e^{- \beta (H^0_L + H^0_R)/4} \left(\opo_{3,L} - \au \opo^{ \dagger}_{3,R} \au^{-1} \right) e^{ \beta (H^0_L + H^0_R)/4}
\ee
\noindent and $\opo_3 \equiv [\opo_1,\opo_2]$.  
This property can be shown by considering $[d_1, d_2]\ketT=0$ and simplifying the expression while using $[\opo_L, \opo_R] = 0$ and properties of Hermitian conjugation. 

Thus if one has a set  $\mathcal{A}$ such that the elements of $\mathcal{A}$ generate all the operators in the QFT by commutation algebra, then the TFD hamiltonian need only be defined as,
\begin{equation}
 \HTFD = \sum_{ i \in \mathcal{A}} \,   \, c_i\, d_i^\dagger \, d_i \,.
\end{equation}
$\HTFD$ is manifestly positive-definite, being a sum of positive-definite terms.

In principle, this hamiltonian could have more than one ground state. We return to this later when we calculate the gap for Quantum Field Theories. For finite dimensional Hilbert spaces it is easy to prove.

\paragraph{Uniqueness of the ground state} \hfill

One can easily prove that the ground state of this Hamiltonian is unique. The trick consists in linearly mapping the Hilbert space of the double theory to the space of operators of the single sided left theory. Provided a choice of an anti-unitary operator $\Theta$ we can define a linear map $\mathcal{M}$ as
\be
 \mathcal{M} : \ket{n}_{L} \otimes \ket{{m^*}}_{R}  \rightarrow \ket{n} \otimes \bra{{m}}\, .
 \ee
The problem of finding the original ground state has now mapped to that of finding the set of operators that lie in the kernel of the super-operators related through the linear map $\mathcal{M}$ to (\ref{eq:general_d_def})
\be
\mathcal{D}_i = [\mathcal{O}_i, \cdot]_\beta 
\ee
\noindent for all $i \in \mathcal{A}$. The $\beta$-commutator above is defined as:
\be
[\mathcal{O}, \mathcal{Q} ]_\beta=e^{- \beta H^0/4} \mathcal{O} e^{+\beta H^0/4} \mathcal{Q} - \mathcal{Q}  e^{+ \beta H^0/4}\mathcal{O} e^{-\beta H^0/4} 
\ee

Now, the only operator that $\beta$-commutes with all operators in $\mathcal{A}$ is the $\beta$-identity $\mathcal{I}_\beta = e^{- \beta H^0/2}$. Under the inverse map $\mathcal{M}^{-1}$, this operator corresponds to the thermofield double state. Therefore, $\mathcal{I}_\beta$ is the unique ground state of the super-Hamiltonian.
\be
\mathcal{H}_{TFD} =  \sum_{ i \in \mathcal{A}} \,   \, c_i\,  [ \mathcal{O}_i^\dag ,[\mathcal{O}_i, \cdot]_\beta ]_{-\beta}
\ee
As before, $c_i$ is a set of positive numbers.  It is easy to check that this super-Hamiltonian has a positive semi-definite spectrum as the expectation value in any state $\mathcal{Q}$ is given by:
\be
\langle \mathcal{H}_{TFD} \rangle_\mathcal{Q} =\sum_{ i \in \mathcal{A}} \,   \, c_i\,  \tr \,\{ \mathcal{Q}^\dag [ \mathcal{O}_i^\dag ,[\mathcal{O}_i, \mathcal{Q}]_\beta ]_{-\beta} \}=\sum_{ i \in \mathcal{A}} \,   \, c_i\,  \tr \, \{[\mathcal{O}_i, \mathcal{Q}]_\beta^\dag[\mathcal{O}_i, \mathcal{Q}]_\beta\} \geq 0
\ee
and has a unique ground state with $\mathcal{H}_{TFD}=0$ for the state $\mathcal{I}_\beta$.

This proof is valid in QFT  provided we can regularize the sum over all $i$'s in $\mathcal{A}$ in the case of an infinite dimensional Hilbert space.

\paragraph{Conformal Theories} \hfill

The special case where the quantum systems have conformal invariance is particularly interesting for holography. For now we do not specify whether the theory is conformal quantum mechanics or conformal field theory. Consider the following form of the TFD Hamiltonian
\begin{align}
\begin{split}
\label{eq:desc_htfd}
\HTFD &= \sum_\alpha \lambda_\alpha \,  d_\alpha^\dagger \, d_\alpha +  \sum_{ i}    \, c_i \, d_i^\dagger \, d_i \, , \\
d_\alpha &\equiv e^{- \beta (H^0_L + H^0_R)/4} \left(J_\alpha^L - \au J_\alpha^{R, \dagger} \au^{-1} \right) e^{ \beta (H^0_L + H^0_R)/4} 
\end{split}
\end{align}
where $J_\alpha^{L,R}$ are the generators of the conformal algebra in the left and right theories, and $d_i$ is as defined in \eqref{eq:general_d_def} for a \textit{primary} operator $\opo_i$. We can show that there is no need to separately include $d$ operators constructed from the descendants in the TFD Hamiltonian:
\be\label{lemma1} 
d_{\opo_i} \ketT =0  \, \implies \, d_{J(\opo_i)} \ketT =0 
\ee
where $d_{J(\opo_i)}$ denotes the $d$ operator constructed from a descendant of $\opo_i$. This property follows immediately upon using the commutator property.


Notice that in this case the terms proportional to $\lambda_\alpha$ make the Hamiltonian non-local at all scales.

\paragraph{Ambiguity in the TFD Hamiltonian} \hfill
\label{ambiguousH}

We also note that the TFD Hamiltonian is ambiguous. In fact, all the operators of the following form also annihilate the TFD state and hence in principle can compose the TFD Hamiltonian,
\begin{itemize}
\item $d^{(1)} \equiv   e^{-\frac{\beta}{4} (H_L+H_R)} \, \bigg( O_L - \au O_R^\dagger \au^{-1}\bigg) \, \, e^{\frac{\beta}{4} \, (H_L+H_R)}  $
\item $d^{(2)} \equiv O_L - e^{-\beta H_R/2} \, \au \, O^{ \dagger}_R \, \au^{-1} \, e^{\beta H_R/2}$
\item $d^{(3)} \equiv e^{-\beta H_L/2} \,\au\, O_{L} \, \au^{-1} \, e^{\beta H_L/2} - O_R^\dagger$ 
\item $d^{(4)} \equiv   e^{-\beta H_L/4} \, O_L \, e^{\beta H_L/2} - \au O_R^\dagger \, \au^{-1} \, e^{\beta H_R/4} $
\end{itemize}
We will primarily use the first and the second of these.

\paragraph{Nonlocality in the TFD hamiltonian} \hfill

In the context of quantum field theory, it is natural to ask  how local the interactions are. In other words, do they only couple operators at the same spacetime point in the two copies, or is the coupling non-local?  We will address this more fully in the examples, but we can give a quick answer now.

Looking at the operator $d^{(1)}$, we can take $\opo_L$ to be a local operator. We see that this is coupled to an operator in the right theory that is evolved by $\beta/2$ in Euclidean time. Therefore, we expect the right operator to have non-locality roughly on the temperature scale $\beta$. This statement is not precise in general, because evolution in Euclidean time is not contained in any lightcone, so we will calculate the scale of non-locality explicitly in examples. 

Another argument for nonlocality on scale $\beta$ is that  creating the TFD state from two identical QFTs requires entangling them. In many cases, the entanglement between the two systems extends a distance $\beta$ in space.

Given two quantum systems in the lab, we can connect wires coupling nearby points in the two theories. The speed of light in the lab may be much faster than the speed of light in the QFT's, so there is no obstacle to introducing interactions at spacelike separation.

However, note that the operators in the simple Hamiltonian \eqref{eq:simpleh} are local on the scale $\beta$. They are smeared over some short distance $\frac{1}{\sigma_E}$ to regulate their UV behavior, so this yields an Effective Field Theory whenever $\beta \sigma_E \gg 1$. This suggests that the ground state of an approximately local Hamiltonian is close to the TFD state. We will discuss in detail the overlap between the two in Section \ref{approxH}.

\section{Examples}
\label{sec-ex}

In this section we illustrate the general construction above with some concrete, albeit simple examples.  

\subsection{Simple Harmonic Oscillator}
We begin with the harmonic oscillator. We will take the anti-unitary operator in this case to be time reversal. One could also choose $PT$; this would yield a different TFD state that is related to the one we construct here by flipping the axis of one system.

\paragraph{Exact TFD Hamiltonian.}
We first construct our exact TFD Hamiltonian. A convenient choice of annihilation operators is
\begin{align}
\label{ann1}
d_{1}&= a_{L} - e^{-\beta w/2} a_{R}^{\dagger}  \, , \quad d_{2} =  a_{R} - e^{-\beta w/2} a_{L}^{\dagger}  \, .
\end{align}
The Hamiltonian becomes
\beq
\HTFD = E_0 \left( d^\dagger_1 d_1 + d^\dagger_2 d_2 \right) \, ,
\eeq
where $E_0$ is an arbitrary constant. We have chosen the relative coefficient between the first and the second term above to be one. This is the unique quadratic Hamiltonian that respects the symmetry under the exchange of left and right oscillators. 
Collecting terms and dropping a constant shift gives
\beq
\label{htfd2}
\HTFD &=&E_0 (1 + e^{-\beta w}) \left(a_{L}^{\dagger}a_{L} +  a_{R}^{\dagger}a_{R} \right) - 2 E_0 e^{-\beta w/2} \left(a_{L}^{\dagger} a_{R}^{\dagger} +a_{L}a_{R}\right) \, .
\eeq
This can be diagonalized by defining
\beq
a_L \equiv (a + b)/\sqrt{2} \, , \quad a_R \equiv (a-b)/\sqrt{2} \, ,
\eeq
so that $a, a^\dagger $ and $b, b^\dagger$ have the canonical commutators and commute with each other. Then the Hamiltonian becomes
\beq
\label{eq:shoham}
\HTFD = E_0 (1 + e^{-\beta w}) (a^\dagger a + b^\dagger b) - E_0 e^{-\beta w/2} (a^\dagger a^\dagger + a a - b^\dagger b^\dagger - b b) \, .
\eeq

We now have two decoupled systems, each with Hamiltonian of the form
\beq
H = B a^\dagger a +  D [a^2 + (a^\dagger)^2] \, .
\eeq
The spectrum can be calculated by doing a Bogoliubov transformation. The spectrum is that of a harmonic oscillator with a frequency given by
\beq
w' = \sqrt{B^2 -4  D^2}=E_0 (1 - e^{-\beta w}) \, .
\eeq
Looking back at our full TFD Hamiltonian, we see the spectrum is that of two decoupled harmonic oscillators with the same energy spacing. The gap is
\beq
{\rm Gap} = E_0 (1 - e^{-\beta w}) \, .
\label{exactshogap}
\eeq
The high temperature limit of the term in parentheses is $\beta w$, so we need to have our overall constant $E_0$ scale at least like
\begin{equation}
\label{eq:E0T}
E_0 \sim T
\end{equation}
at high temperatures to maintain a finite gap. This is a reasonable requirement. Note that by making different choices for how the overall scale in the Hamiltonian scales with temperature, we can make the gap scale in any way we like. We will find that our simple Hamiltonian has less freedom.

\paragraph{Simple Hamiltonian.}
We can also try out our simple Hamiltonian \eqref{eq:simpleh} for the harmonic oscillator. There is no guarantee that this will give even approximately the correct ground state since we only claim it works in systems satisfying ETH, but we will try anyway. We take 
\beq
H_S = H_L^0 + H_R^0 + c_1 w^2 (x_L - x_R)^2 + c_2 (p_L+p_R)^2 \, .
\eeq
Note that the relative sign is different in the momentum coupling due to conjugation by the time reversal operator. We will tune the constants $c_1$ and $c_2$ to try to get the TFD state as the ground state. We have defined them so that the $c_i$ are dimensionless. If we choose 
\begin{equation}
c_1 = c_2 = C/2 \, ,
\end{equation}
the interaction term becomes (up to a constant shift)
\beq
C w\left[ a^\dagger_L a_L + a^\dagger_R a_R - a_L a_R - a^\dagger_L a^\dagger_R \right] \, ,
\eeq
so that the full Hamiltonian is
\beq
H_S = w \big(1+C \big) (a^\dagger_L a_L + a^\dagger_R a_R) - w C (a_L a_R + a^\dagger_L a^\dagger_R) \, .
\eeq
This is precisely of the same form as the exact TFD Hamiltonian from equation \ref{htfd2}! We were lucky in this case because everything is quadratic. 

Matching parameters, we can relate the interaction coefficient $C$ to the temperature 
\begin{equation}
\label{eq:simpleSHOC}
C = {1 \over 2 \sinh^2(\beta w/4)} \, ,
\end{equation}
indicating that the simple Hamiltonian gives a TFD state with temperature that ranges from $T=0$ when $C=0$ up to $T=\infty$ at $C=\infty$.

The gap can be found be relating $C$ to $E_0$ and is given by
\be
\label{eq:simplegapSHO}
{\rm Gap} = w \, \coth \left(\beta w \over 4\right) \, .
\ee
Note that in this simple Hamiltonian we do not have the freedom to choose the gap. The temperature dependence of the gap is nice: the gap is given by the frequency of the oscillator at low temperature, and by the temperature at high temperature.

\subsection{Free Fermion}
\label{fermi-osc}
We may repeat the steps above for fermions, though we must be careful about orderings.  For fermions, our conventions are
\begin{align}
\begin{split}
\{ a_{L,R}, a_{L,R}^{\dagger}\} = 1,\qquad \{a_{L,R},a_{R,L}^{\dagger}\} = \{a_{L,R},a_{R,L}\}= \{a^{\dagger}_{L,R},a^{\dagger}_{R,L}\}=0 \, .
\end{split}
\end{align}
Note that the Hilbert space of the each fermionic oscillator is finite dimensional, in fact, spanned by two independent states. When we write the vacuum of the doubled theory, we specifically have the following ordering in mind $|0,0\rangle = |0\rangle_{L}\otimes |0\rangle_{R}$.  The excited state is then
\begin{equation}
a_{R}^{\dagger} a_{L}^{\dagger} |0,0\rangle \equiv |1,1\rangle \, .
\end{equation}
The anti-unitary operator $\Theta$ from equation \eqref{eq:general_d_def} acts as follows
\begin{align}
\begin{split}
\Theta \, a_{L,R}^\dagger \, \Theta^{-1} = -a_{L,R}^\dagger \, .
\end{split}
\end{align}
Then, using this and keeping track of orderings, the thermofield double state becomes
\begin{align}
\begin{split}
\label{fermid}
\ketT &= {\rm exp} \big(e^{-\beta w/2} a_{R}^{\dagger} a_{L}^{\dagger} \big) \ket{0,0}  = \ket{00} + e^{-\beta w/2} \, \ket{11} \, .\\
\end{split}
\end{align}
It is annihilated by 
\begin{align}
\begin{split}
d_L &= a_L + e^{-\beta w/2} a_R^{\dagger} \, , \quad d_R = a_R - e^{-\beta w/2} a_L^{\dagger}  \, .
\end{split}
\end{align}
Then the exact TFD Hamiltonian with the TFD state as the ground state can be shown to be,
\begin{align}
 \begin{split}
\HTFD &= E_0 \, \bigg( 1- e^{-\beta w} \bigg) \, (a_L^\dagger \, a_L + a_R^\dagger \, a_R ) + 2 E_0 e^{-\beta w/2} \, (a_L^\dagger \, a_R^\dagger - a_L \, a_R)  \, .
 \end{split}
\end{align}
We can now ask for the gap of this exact Hamiltonian. Since the form of the Hamiltonian is similar to equation \eqref{eq:shoham}, the gap can be calculated in a similar way. It becomes
\begin{equation}
\label{eq:foscgap}
\text{Gap} = E_0 (1+ e^{-\beta w})  \, .
\end{equation}
Comparing this gap to the one in equation (\ref{exactshogap}), we see that bosons and fermions behave very differently at low energies.

\subsection{Free Quantum Field Theory}

We would also like to analyze a simple quantum field theory example in order to diagnose locality. We will analyze the free massless scalar in $3+1$ dimensions for simplicity.  This is of course just a bunch of harmonic oscillators. 
(The $1+1$ case has IR divergences that are special to that case, so we work in higher dimensions.) 

\paragraph{Exact TFD Hamiltonian}
By using the same approach as the harmonic oscillator example for each momentum mode, the exact TFD Hamiltonian becomes,
\begin{align}
\begin{split}
\HTFD &= \int d^3k \, E(k) \, (1 + e^{-\beta \omega_k}) \, \bigg((a_k^L)^\dagger a_k^L + (a_k^R)^\dagger a_k^R \bigg) \\
&\qquad   - 2 E(k) \,  e^{- \beta \omega_k/2} \,  \bigg( (a_k^L)^\dagger (a_{-k}^R)^\dagger + a_k^L a_{-k}^R \bigg) \, .
\end{split}
\end{align}
We are free to choose $E(k)$ to be any positive function we like. We would like to go to position space to diagnose locality. For this we use, 
\be
a_k = \int d^3x \,  e^{-i k x} \,  \left[ \sqrt{\frac{\omega_k}{2}} \phi(x) + {i \over \sqrt{2 \, \omega_k}} \pi(x) \right] \, .
\ee
Then, up to additive constant factors which will not be important for further discussion, the Hamiltonian becomes
\begin{align}
\begin{split}
H &= {1 \over 2} \int d^3x \, d^3 y \,  \bigg[ f(x-y) \, \bigg(\pi_L(x) \pi_L(y) + \pi_R(x) \pi_R(y) \bigg) \\
&\qquad  + g(x-y) \bigg(\phi_L(x) \phi_L(y) + \phi_R(x) \phi_R(y) \bigg) \bigg]  \\ 
&\quad + {1 \over 2} \int d^3x \, d^3 y  \, \bigg[ h(x-y) \,  \pi_L(x) \pi_R(y)   + k(x-y) \,  \phi_L(x) \phi_R(y) \bigg]  \, .
\end{split}
\end{align}
Here the first line contains terms that do not couple the two theories, while the second line contains coupling terms. This Hamiltonian is bi-local, with the scale of nonlocality set by the four functions $f, g, h, k$. These functions are all determined by our choice of $E(k)$ via the definitions
\begin{align}
 \begin{split} \label{nonlocfun}
f(x)  &= \int d^3k \, e^{ikx} \,  {E(k) \, (1 + e^{-\beta \omega_k}) \over \omega_k}  \, , \\
g(x) &= \int d^3k \, e^{ikx} \,  E(k) \, \omega_k \,  (1 + e^{-\beta \omega_k}) \, , \\
h(x)  &=  2 \int d^3k \,  e^{ikx} \,  {E(k) \, e^{-\beta \omega_k/2} \over \omega_k}  \, , \\
k(x)  &=-2 \int d^3k \, e^{ikx} \, E(k) \, \omega_k \,  e^{-\beta \omega_k/2}  \, .
 \end{split}
\end{align}
It is tempting to choose $E(k)$ so that the non-interacting terms take their canonical local form. This choice corresponds to  
\be
E(k) (1+ e^{-\beta \omega_k}) = \omega_k \ .
\label{ekcan}
\ee
However, we do not want to make this choice because the gap for each mode is given by our harmonic oscillator formula (\ref{exactshogap}),
\beq
{\rm Gap}(k) = E(k) (1 - e^{-\beta \omega_k}) \, .
\label{gapkeqn}
\eeq
If we choose $E(k)$ according to (\ref{ekcan}), we would have 
\beq
\label{cangapeq}
{\rm Gap}(k) = \omega_k \tanh (\beta \omega_k/2) \, .
\eeq
Note that the appearance of $tanh$ here is not inconsistent with the appearance of $coth$ in equation \eqref{eq:simplegapSHO}. These are gaps of two different Hamiltonians: equation (\ref{cangapeq}) is that of the exact TFD Hamiltonian while equation \eqref{eq:simplegapSHO} is that of a Simple Hamiltonian. 


Further, at small $\omega$, the gap in equation (\ref{cangapeq}) becomes ${\rm Gap} \sim \beta \omega_k^2$, so if $\omega$ becomes very small, the gap is very very small if we insist on the canonical choice for the non-interacting terms. In fact, it is not possible in this case to confine the nonlocality to the thermal scale while also avoiding a small gap. The gap equation at small $k$ becomes
\be
{\rm Gap}(k) \approx E(k) \beta \omega_k \, .
\ee
For a massless field $\omega_k = |k|$. Thus if we want the gap to remain finite as $k \to 0$, we need $E(k)$ to diverge at least as $E(k) \sim 1/k$. However, this behavior leads to non-locality at large scales. Roughly, this is because the low $k$ behavior corresponds to long distances. More precisely, if we look for example at the function $f(x)$ defined above in equation (\ref{nonlocfun}), we see that it is the Fourier transform of a function that diverges as at least $1/k^2$ at small $k$, since we want $E(k) \sim 1/k$. In general, the Fourier transform of a function that is non-analytic at $k=0$ cannot fall off exponentially at large $x$ , as is true in this particular case since the function $f(x) \sim x$ as $k \to 0$. 

Therefore, in this example we have to choose between a small gap and an approximately local Hamiltonian. We will describe the case of an approximately local Hamiltonian. That is, we take the function $E(k)$ defined by equation (\ref{ekcan}). Then the non-interacting terms become completely local. This becomes manifest when we calculate the functions appearing in the Hamiltonian, obtaining (up to constants) 
\begin{align}
 \begin{split} \label{finalheq}  
f(x) &= \delta^3(x)  \, , \\
g(x) &= -\nabla^2 \delta^3(x)   \, , \\
h(x) &= {1 \over 8 \beta^2 |x|} {\sinh\left(\pi |x| \over 2 \beta \right)\over \cosh^2\left(\pi |x| \over 2 \beta \right)}  \, , \\
k(x) &= \nabla^2 h(x) \, .
 \end{split}
\end{align}
Here, we have included the important property that $h$ and $f$ are equal at long wavelengths. Collecting everything, the TFD Hamiltonian becomes (up to constant additive factors)
\be
H_{\rm TFD} = H^0_L + H^0_R +{1 \over 2} \int d^3 x \, d^3 y \, h(x-y) \, \bigg[ \pi_L(x) \, \pi_R(y) - \nabla \phi_L(x) \cdot \nabla \phi_R(y) \bigg] \, .
\ee
The scale of nonlocality is set by the function $h$ in (\ref{finalheq}), so it is nonlocal on the thermal scale. The gap can be calculated from equations (\ref{gapkeqn}) and (\ref{ekcan}), giving at small $k$
\be
{\rm Gap} \sim \beta \omega_k^2 \, .
\ee
This is very small at high temperature. We believe that this small gap is an artifact of working in the free theory. We will argue later that interacting theories have a gap of order the temperature. This is reminiscent of the appearance of  thermal masses in finite temperature QFT at non-zero coupling.

\subsection{Free Fermion Field Theory}

We now make a brief comment about free fermion field theory. We start by thinking of the field theory as a collection of decoupled fermion oscillators. The case of free fermionic oscillator was worked out in detail in Subsection \ref{fermi-osc}. Using the results there we can immediately write down the gap of the exact TFD Hamiltonian as
\begin{equation}
\label{eq:fermiFTgap}
\text{Gap}(k) = E(k) (1+ e^{-\beta \omega_k}) \, .
\end{equation}
If we choose to make the non-interacting terms canonical, corresponding to the choice
\be
E(k) = {\omega_k \over 1 - e^{-\beta \omega_k}} \, ,
\ee
the gap becomes
\be
{\rm Gap}(k) = \omega \coth(\beta \omega_k/2 ) \, .
\ee
This is quite different from the free boson field theory gap in equation (\ref{cangapeq}), and disagrees with the claim in \cite{Swingle:2016foj} that the TFD Hamiltonian for free fermions and free bosons behave similarly.

Therefore, for free fermionic fields, a finite range interaction $can$ give the TFD as a ground state while maintaining a gap of order the temperature, unlike the free bosonic case. This is analogous to the finite temperature behavior of free fermions: due to the anti-periodic boundary conditions fermions have no zero mode on the thermal circle, so their finite temperature correlation function is exponentially suppressed with length scale set by the temperature, corresponding to all light modes acquiring a mass of order the temperature.  

On the other hand, free bosonic theories $do$ have a zero mode on the thermal circle, leading to power law correlation functions, which implies that some modes remain much lighter than the thermal scale. Therefore, the different gaps we find for free fermions and free bosons are surprising, but the same as known finite temperature physics. We expect interactions to modify the unusual finite temperature behavior of free bosonic fields. We return to this in the following section.

\subsection{Ising Conformal Field Theory}
\label{ising_model}

It is illustrative to consider the critical ($\beta J = 1/4$) Ising model in the language of conformal field theory. 
The central charge of this theory is $c=\frac{1}{2}$ and there are three conformal primaries:
\begin{center}
\begin{tabular}{ |c|c|c|}
\hline
Operator & Symbol & Conformal dimension $h$  \\
\hline \hline
Identity & $I$ & $h=0$ \\ 
\hline
Spin & $\sigma(z,\bar{z})$ & $h=\frac{1}{16}$ \\ 
\hline
Energy & $\epsilon(z,\bar{z})$ & $h=\frac{1}{2}$ \\ 
\hline
\end{tabular}
\end{center}
\vspace{4mm}
The operator product expansions are:
\begin{align}
\begin{split}
\epsilon(z,\overline{z})\epsilon(w,\overline{w})&= \frac{1}{|z-w|^{2}} \, , \\ 
\sigma(z,\overline{z}) \sigma(w,\overline{w}) &= \frac{1}{|z-w|^{1/4}} + \frac{1}{2} |z-w|^{3/4} \epsilon(w) \, , \\
\epsilon(z,\overline{z}) \sigma(w,\overline{w}) &= \frac{1}{2|z-w|} \sigma(w) \, .
\end{split}
\end{align}
For each operator, we define states and their conjugates via
\begin{align}
\begin{split}
\label{cftdefs}
\ket{\phi_{in}} \equiv \lim_{z,\overline{z}\rightarrow 0} &\phi(z,\overline{z}) \ket{0} , \qquad 
\bra{\phi_{out}} = \ket{\phi_{in}}^{\dagger} \, , \\ 
\phi(z,\overline{z})^{\dagger} &= \overline{z}^{-2 h}z^{-2 \overline{h}} \phi(1/\overline{z}, 1/z) \, .
\end{split}
\end{align}
Using these equations we can convert the operators into $3\times 3$ matrices acting on the basis of states formed from the primaries.  Ordering the basis vectors as $\ket{1} = \ket{\mathbb{I}},\,\ket{2} = \ket{\sigma}$ and $\ket{3} = \ket{\epsilon}$, we have
\begin{equation}
\bra{i} \epsilon(z,\overline{z}) \ket{j} = \left(\begin{array}{ccc} 0 & 0 & \frac{1}{| z|^{2}}  \\ 0 & \frac{1}{2|z|} & 0 \\ 1 & 0 & 0\end{array}\right)  , \qquad 
\bra{i} \sigma(z,\overline{z}) \ket{j} = \left( \begin{array}{ccc} 0 & \frac{1}{|z|^{1/4}} & 0 \\ 1 & 0 & \frac{1}{2 |z|}  \\ 0 & \frac{|z|^{3/4}}{2} &0 \end{array} \right) \, .
\end{equation}
Using the general construction of section \ref{sec_general_constr} we can form operators which annihilate the thermofield double
\begin{align}
\begin{split}
d_{\epsilon} &= \epsilon_{L}(z) - e^{-\beta H/2} \epsilon_{R}^{\dagger}(z) e^{\beta H/2} \, , \\ 
d_{\sigma} &= \sigma_{L}(z) - e^{-\beta H/2} \sigma_{R}^{\dagger}(z) e^{\beta H/2} \, , \\ 
\end{split}
\end{align}
Now it is a simple matter to find the matrices representing $d_{\epsilon}$ and $d_{\sigma}$.  Recall that these are acting on a 9 dimensional Hilbert space (i.e., tensor product of two sides).  We should thus consider the $9\times 9$ matrix:
\begin{align}
\left(d_{\sigma}(z)\right)_{i i', jj'}&= \sigma_{i i'}(z)\otimes \mathbb{I}_{jj'} - e^{\beta /16} \mathbb{I}_{ii'} \otimes \sigma^{\star}(e^{\beta/2}z)_{j'j} \, ,
\end{align}
and likewise for $d_{\epsilon}(z)$.  It is now straightforward to check that $d_{\sigma}$ and $d_{\epsilon}$ annihilate the thermofield double for any value of $z$.  Moreoever, in the nine-dimensional tensor product space $\mathcal{H}_{L}\otimes \mathcal{H}_{R}$ (truncated to primaries) one can easily check that the thermofield double is the unique  state annihilated by both $d_{\epsilon(z)}$ and $d_{\sigma(z)}$ for all $z$. Therefore, it is natural to propose the Hamiltonian
\begin{equation} 
\label{eq:isingHfinal}
\HTFD = \int d\theta \left(c_{1} d^{\dagger}_{\epsilon} d_{\epsilon} + c_{2} d^{\dagger}_{\sigma} d_{\sigma}\right)~.
\end{equation}
Note that  this can be put into the form \eqref{eq:desc_htfd} if we identify the sum over $i$ with the integral over $\theta$.

This Hamiltonian $may$ have the TFD as the unique ground state, but we have only established this in the space of primaries. It may be necessary to add terms quadratic in the conformal generators, as in \eqref{eq:isingHfinal}. As mentioned near that equation, these terms may introduce UV issues. It would be interesting to understand the construction further in tractable examples such as the Ising model, but we leave this for future work.

\section{Effective Field Theory}
\label{EFT}

The exact TFD Hamiltonian has two annoying features in general: it is nonlocal, and it is complicated. This motivates us to seek a simpler approximate form for the Hamiltonian. This will turn out to be possible, but at the cost of working in an effective field theory whose UV cutoff is the temperature scale. 

We start with the symmetric form of our annihilation operator
\begin{equation}
d_O \equiv   e^{-\frac{\beta}{4} (H_L+H_R)} \, \bigg( O_L - \au O_R^\dagger \au^{-1}\bigg) \, \, e^{\frac{\beta}{4} \, (H_L+H_R)}  \, .
\end{equation}
For convenience, let us define the following shorthands,
\begin{align}
\opo &\equiv O_L - \au O_R^\dagger \au^{-1}  \, , \quad \mathcal{H} \equiv H_L+H_R \, .
\end{align}
Then we can make the appearance of higher-dimension operators manifest in $d_O$ by using the BCH formula to expand it in powers of $\beta$. We obtain,
\be
d_O = \opo - {\beta \over 4} [\mathcal{H},  \opo] + {\beta^2 \over 32} [\mathcal{H}, [\mathcal{H},  \opo]] + \cdots \, .
\ee
The contribution from $d_O$ to the TFD Hamiltonian is
\begin{align}
 \begin{split}
H_\opo &=  \opo^\dagger \opo + {\beta \over 4} \, \bigg( [\mathcal{H}, \opo^\dagger] \opo - \opo^\dagger [\mathcal{H}, \opo] \bigg)  \\
 &\quad + {\beta^2 \over  32} \, \bigg( [\mathcal{H}, [\mathcal{H}, \opo^\dagger]] \opo + \opo^\dagger[\mathcal{H}, [\mathcal{H}, \opo]] -[\mathcal{H}, \opo^\dagger] [\mathcal{H}, \opo] \bigg)  \, .
 \end{split}
\end{align}
If we now take $O_L$ to be a local operator, this is an expansion in local operators. Due to additional commutators with the Hamiltonian, the higher powers of $\beta$ multiply higher dimension operators. Therefore, this Hamiltonian must be interpreted in a theory with a UV cutoff below the temperature scale. We will illustrate the use of this formula using an example. 

\subsection*{Free Field Theory Example} 

It is straightforward to apply the previous results to free quantum field theories.  Let us consider a massive bosonic theory, although a similar procedure will hold for a fermionic theory. The original Hamiltonian in position space may be written as,
\be
H = {1 \over 2} \int d^3x \left( \pi(x)^{2} + \w^{2} \phi(x)^{2}\right) \, ,
\ee
where $\w^{2}$ is shorthand for the operator 
\be
\w^2 \equiv -\nabla^{2} + m^{2} \, .
\ee  
In this case if we take  $O_{L,R} = \phi_{L,R}(x)$ , we can calculate the full operator appearing in the TFD Hamiltonian,
\be
e^{-\beta \mathcal{H}/4} \, \opo \, e^{\beta \mathcal{H}/4} \, .
\ee
where $\mathcal{H}=H_L+H_R$. To avoid factors of $4$ we define
\be
\gamma \equiv \beta/4 \, .
\ee
Again using the BCH formula and the canonical commutator $[\phi(x),\pi(y)] = i \delta(x-y)$, we can show that for each of the field theory,
\be
e^{-\gamma H}\phi(x) e^{\gamma H} = \cosh(\gamma \w ) \phi(x) + \frac{i}{\w} \sinh(\gamma \w)  \pi(x) \, .
\ee
Plugging this expression into the formula for $d_{\phi(x)}$ we get
\be
d_{\phi(x)} = E_0 \left[ \cosh(\gamma \w)\phi_L(x) + i {\sinh(\gamma \w) \over \w} \pi_L(x) - \cosh(\gamma \w)\phi_R(x) - i {\sinh(\gamma \w) \over \w} \pi_R(x) \right] \, ,
\ee
where the overall constant $E_0$ has units of energy. 

Now we construct the TFD Hamiltonian. Since we are working with local operators $\phi(x)$, the natural TFD Hamiltonian is an integral
\be
H_{\rm TFD} = \int d^3x \,  d^\dagger_{\phi(x)} d_{\phi(x)} \, .
\ee
Explicitly, the full Hamiltonian is a bit of a mess,
\begin{align}
\begin{split}
H_{\rm TFD} &= E_0^2 \int d^3x \, \left[   \cosh(\gamma \w)(\phi_L - \phi_R) - i  {\sinh(\gamma \w_L) \over \w_L} (\pi_L- \pi_R) \right] \times  \\
 &\qquad \left[   \cosh(\gamma \w)(\phi_L - \phi_R) + i  {\sinh(\gamma \w_L) \over \w_L} (\pi_L- \pi_R) \right] \, .
\end{split}
 \end{align}

Note that terms like $\cosh(\gamma \w)$ tell us that this Hamiltonian is only local on the thermal scale, because expanding out the $\cosh$ gives higher powers of the momentum $\beta^2 \nabla^2$.   This might lead one to suspect that the $T\rightarrow \infty$ limit is completely local.  Indeed,  in this limit, if we expand for small $\beta$ and keep operators up to dimension 2 the TFD Hamiltonian just approaches
\be
H_{\rm TFD}' =E_0^2 \int d^3x \left[ (\phi_L - \phi_R)^2 + \gamma^2 (\phi_L - \phi_R) \w^2 (\phi_L - \phi_R) + \gamma^2(\pi_L - \pi_R)^2 \right] \, .
\ee
It is natural to choose the overall dimensionful constant $E_0$ to be set by the temperature scale, $E_0 =\frac{\gamma^{-1}}{\sqrt{2}}$, giving
\be
H_{\rm TFD} \approx {1 \over 2} \int d^3x \, \left[  (\phi_L - \phi_R) \w^2 (\phi_L - \phi_R) + (\pi_L - \pi_R)^2  + 8 T^2  (\phi_L - \phi_R)^2\right] \, .
\ee
This Hamiltonian is weird because it only has a kinetic term for one linear combination of the fields. It is natural to add a second term to the Hamiltonian where we start from the operator $\pi(x)$ instead of $\phi(x)$. This operator is higher dimension so we only need to expand to leading order in $\beta$. Due to the conjugation by the anti-unitary operator, this gives a term
\be
d_\pi^\dagger d_\pi =  (\pi_L + \pi_R)^2 + \dots \, .
\ee
We can also add a term from $\opo = \partial \phi$ which contributes
\be
d_{\partial_i \phi}^\dagger d_{\partial_i \phi} = (\partial_i \phi_L - \partial_i \phi_R)^2 + \dots \, .
\ee
Combining all three of these terms with arbitrary positive coefficients $c_i$  gives the full Hamiltonian at quadratic order in the fields up to the operator dimension we are working,
\begin{align}
 \begin{split}
H_{\rm TFD} &\approx {1 \over 2} \int d^3x \, \bigg[ c_1(\pi_L - \pi_R)^2 + c_2(\pi_L + \pi_R)^2 + \\
&\qquad  (\phi_L - \phi_R) (c_1 \w^2 -c_3 \nabla^2 + 8 c_1 T^2) (\phi_L - \phi_R) \bigg]  \, ,
 \end{split}
\end{align}
where the dimensionless constants $c_i$ can be freely chosen. The most notable aspect of this Hamiltonian is that while the combination $\phi_L - \phi_R$ gets a mass as well as a gradient term, the combination $\phi_L + \phi_R$ has no potential or gradient term to this order. This term will appear at higher order in the term that begins with $\pi$,
\be
d_\pi^\dagger d_\pi = (\pi_+)^2 + \gamma^2 \pi_+ \w^2 \pi_+ + \gamma^2 (\w^2 \phi_+)^2 + \dots \, ,
\ee
where we have defined $\pi_+ \equiv \pi_R + \pi_L$. The Hamiltonian factorizes into a $\phi_+$ and $\phi_-$ piece,
\begin{align}
 \begin{split}
H_{\rm TFD} &\approx {1 \over 2} \int d^3x \, \bigg[ c_1 \pi_-^2 + \phi_- (c_1 \w^2 -c_3\nabla^2 + 8 c_1 T^2)\phi_- + \\
&\qquad  c_2 \pi_+^2 + {8 c_2 \over T^2}  \pi_+ \w^2 \pi_++  {8 c_2 \over T^2}\left( \w^2 \phi_+ \right)^2 \bigg] \, .
 \end{split}
\end{align}
This confirms what we saw previously in a simpler way: we have one mode with a gap set by the temperature, and a light mode with 
\be
{\rm Gap} \sim c_2{k^2+m^2 \over T} \, .
\ee  
The constant $c_2$ can be freely chosen, but in a massless theory the gap is set by lowest allowed value of $k$; in other words, the gap is set by the IR cutoff of the theory.

It would be interesting to know if there are more general situations where a light mode appears, or if this is simply an artifact of free field theory. Our prejudice is the latter. This is an important question because using our Hamiltonian to cool to the TFD state becomes difficult whenever the gap is small.

As motivation that the small gap is an artifact, consider a $\lambda \phi^4$ interaction in four dimensions. We have
\be
d_\pi = \pi_+ - \gamma \left[ \mathcal{H}, \pi_+\right] + \dots \, ,
\ee
where as above $\mathcal{H} \equiv H_L + H_R$. One of the terms in the commutator gives
\be
d_\pi = \pi_+ + \# i \gamma \lambda \phi_+^3 + \dots \, ,
\ee
which contributes to the Hamiltonian as
\be
H_{\rm TFD} = \int d^3x \,  \left[ \pi_+^2 + i \gamma \lambda [\phi^3(x), \pi(x)]  + \dots \right] \, .
\ee
The commutator gives
\be
[\phi(x)^3, \pi(x)] = - i \delta^3(0) \, .
\ee
Since we are working in effective field theory with cutoff of order temperature, $\delta^3(0)$ should be replaced by $T^3$, so that the TFD Hamiltonian includes the terms
\be
  \int d^3x \, \left[ \pi_+^2 + \# T^2 \lambda \phi_+^2 \right] \subset H_{\rm TFD} \, .
\ee
This is a mass term for the $\phi_+$ mode with 
\be
m^2_+ \sim \lambda T^2 \, .
\ee
This shows that instead of a gap that depends on the IR cutoff, as in the free theory, the $\lambda \phi^4$ theory has a gap set by the temperature scale. As discussed earlier, this phenomenon is not special to our TFD analysis; the same thing happens in analyzing field theory at finite temperature. Free bosonic theories at finite temperature have a small gap that depends on the IR cutoff, while interacting theories (as well as free fermionic theories) have a gap proportional to the temperature.

\section{TFD in Systems Satisfying ETH}
\label{sec-exacth}

In order to construct the TFD in systems that are dual to classical wormholes, we need to move beyond these simple examples.
It turns out that we can find a relatively simple Hamiltonian whose ground state is the thermofield double in any quantum field theory satisfying the eigenvalue thermalization ansatz (ETH). Conformal symmetry is not needed. We will start by discussing the spectrum of the TFD Hamiltonian in an energy window in Section \ref{energy_window} and show that the ground state is the (infinite temperature) TFD state. In Section \ref{full_Hspace} we will extend our analysis to the full Hilbert space and study the finite temperature spectrum of the TFD state. We will show in Section \ref{full_Hspace} that the gap is order one, indicating that the TFD state can be reached in reasonable time.

\subsection{Analysis in an Energy Window}
\label{energy_window}

To get a feel for what happens for an ensemble of operators obeying ETH, consider our TFD hamiltonian in the infinite temperature limit. We include $K$ operators in our Hamiltonian, which is
\begin{equation}
\label{eq:H_window1}
H_{TFD} = \sum_{k=1}^{K} c_k \left(\mathcal{O^\dagger}_{L}^{k} - \mathcal{O}^{k*}_{R}\right)\left(\mathcal{O}_{L}^{k} - \mathcal{O}^{k T}_{R}\right) \, .
\end{equation}
To avoid clutter, we have defined 
\begin{equation}
\opo^T \equiv \au \opo^\dagger \au^{-1} \, ,  \quad \opo^* \equiv (\opo^T)^\dagger \, .
\end{equation}
The notation is natural because in the case that the eigenstates are invariant under $\au$, $\opo^T$ is simply the transpose in the energy basis. We further assume that the eigenvalue thermalization hypothesis (ETH) applies to each copy of the original theory. ETH says that the matrix elements of the operator $\mathcal{O}$ obey
\begin{equation}
\bra{i} \mathcal{O} \ket{j} = \langle \mathcal{O}\rangle_{T} \delta_{ij} + {1 \over \sqrt{\rho(\bar{E})}} \xi_\opo(\bar{E}, \omega) R_{ij} \, ,
\label{ethans}
\end{equation}
where $\bar{E}$ is the average energy of the two states,  $\omega$ is the energy difference, and $\rho(E) \approx \exp(S)$ is the density of states. $R_{ij}$ is a random matrix whose elements have mean zero and unit variance, while $\xi_\opo$ is a smooth function of the energy difference $\omega$ and the average energy $\bar{E}$. 

\begin{figure}[!htb]
\begin{center}
\includegraphics[scale=1]{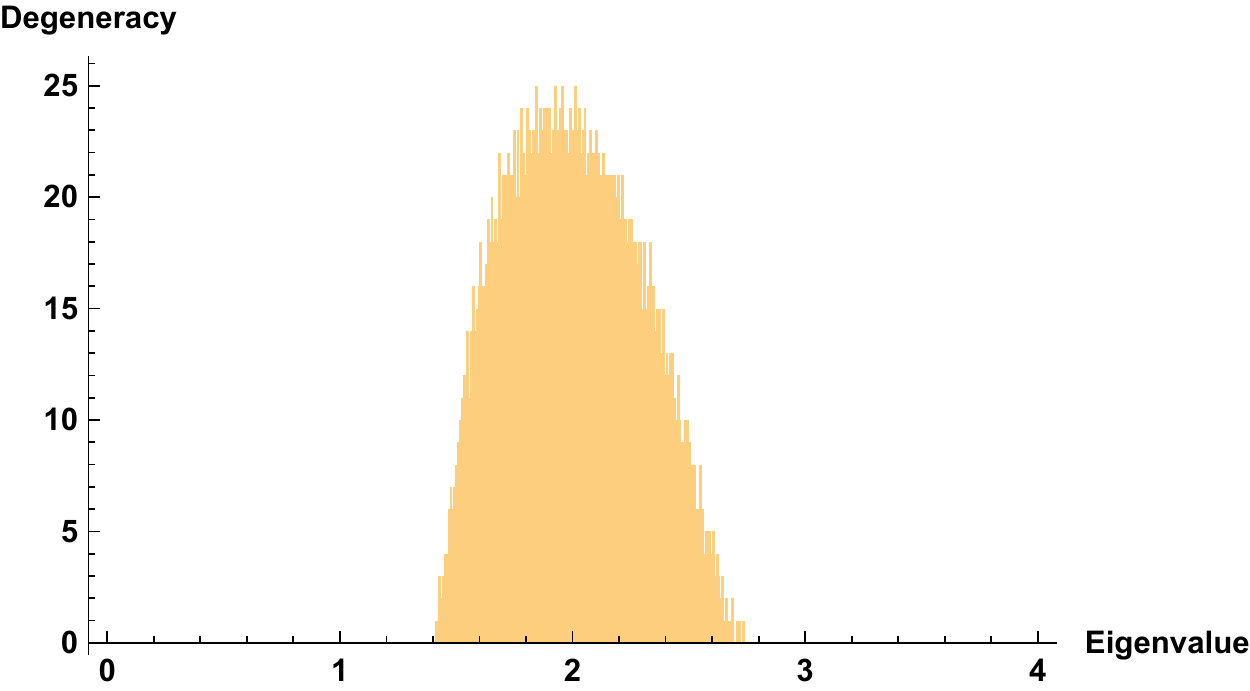}
\end{center}
\caption{Spectrum of $\HTFD$ in energy window. K=N=50.}
\label{fig:EVspectrumLargeK}
\end{figure}

The Hamiltonian is an operator in the doubled Hilbert space. A general matrix element of the Hamiltonian in the basis of energy eigenstates is
\begin{align}
\begin{split}
\label{eq:genHelement}
\bra{a_L i_R} H_{\rm TFD} \ket{b_L j_R} &= \sum_{k=1}^K c_k \left(\bra{a} \opo^{\dagger }_k \opo_k \ket{b} \delta_{ij} + \delta_{ab} \bra{j} \opo^{\dagger }_k \opo_k \ket{i}\right)  \\
 &- \sum_{k=1}^K c_k \left( \bra{a} \opo_k^\dagger \ket{b} \bra{j} \opo_k \ket{i}   + \bra{a} \opo_k \ket{b}^* \bra{i} \opo_k \ket{j} \right) \, .
\end{split}
\end{align}
Note that all matrix elements appearing here are quantities in a \textit{single} copy of the theory. After inserting a complete basis of energy eigenstates in the formula \eqref{eq:genHelement}, it becomes
\begin{align}
 \begin{split}
 \label{eq:H_ele_basis}
  \bra{a i} H_{\rm TFD} \ket{b j} &= \delta_{ij} \sum_{k=1}^K c_k \sum_d  \bra{a} \opo^{\dagger }_k \ket{d}  \bra{d} \opo_k \ket{b} + \delta_{ab} \sum_{k=1}^K  c_k \sum_{\ell}  \bra{j} \opo^{\dagger }_k \ket{\ell}  \bra{\ell} \opo_k \ket{i}  \\
 &\quad - \sum_{k=1}^K c_k \left( \bra{a} \opo_k^\dagger \ket{b} \bra{j} \opo_k \ket{i}   + \bra{a} \opo_k \ket{b}^* \bra{i} \opo_k \ket{j} \right) \, .
 \end{split}
\end{align}
We now specialize our discussion to an energy window with $N$ states, where $N= \exp(S)$. In the window, the energy-dependence of the operators $\opo_k$ can be taken to be constant. The ETH then simplifies to 
\begin{equation}
\label{eq:toy_ETH}
\bra{a} O_k \ket{b} = \frac{1}{\sqrt{N}} R^k_{ab} \, ,
\end{equation}
where $a,b,k=1,2,\cdots , N$ and the matrix element $R_{ab}$ is a real random variable with mean $\tilde{\mu}=0$ and standard deviation $\nu = 1$. It is often taken to be Gaussian. If we use equation \eqref{eq:toy_ETH} to insert the Hamiltonian \eqref{eq:genHelement} in a computer, we can numerically calculate the spectrum of the Hamiltonian. For $N=50$, we plot it in Figure \ref{fig:EVspectrumLargeK}. Although it is not clear from the figure, there is an eigenvalue at zero, since we know that $\HTFD$ annihilates the TFD state.

 We further assume that the operators $\opo_k$ are low-energy and have a soft UV behavior. They can be thought of as having an \textit{effective radius} in energy space and do not connect energy eigenstates separated by distance larger than the width of the window. Then the complete basis inserted in equation \eqref{eq:H_ele_basis} simplifies and only those eigenstates $\ket{d},\ket{\ell}$ give a non-zero contribution which are in the energy window. There are $N$ such states. The simplified ETH in equation \eqref{eq:toy_ETH} then implies that the first two terms in equation \eqref{eq:H_ele_basis} have an explicit $N$ in front of them.

\subsubsection{Large K}
\label{largeK}

Unlike the terms on the first line, those on the second line in equation \eqref{eq:H_ele_basis} are in general not sign definite. However, the off-diagonal terms in the second line are random variables with mean zero and variance one, as per ETH. If $K$ is large, we can use the central limit theorem to conclude that the sum over $K$ of these will give us a random Gaussian variable with mean zero but variance $\sqrt{K}$. In that case, the terms in the second line will be of order $\opo(\sqrt{K})$ in general. 

In this subsection, we will then assume $K \gg 1, K \approx N$ and study the spectrum of the TFD Hamiltonian \eqref{eq:H_ele_basis}. Since the terms in the second line in equation \eqref{eq:H_ele_basis} are of order $\mathcal{O}(\sqrt{N})$, we can ignore the terms in the second line relative to those in the first line. However, there are some off-diagonal terms in the second line that have a definite sign and for which the above argument using central limit theorem does not hold. These are the terms where $a=i$ and $b=j$, so that the last two terms in the Hamiltonian are sums of absolute squares. In this case, the contributions from different operators are of the order $\opo(N)$ and not $\opo(\sqrt{N})$. Keeping these terms, the Hamiltonian becomes
\begin{align}
\begin{split}
\label{eq:Hels_final}
\bra{a i} H_{\rm TFD} \ket{b j} &=\delta_{ab} \delta_{ij} \, \sum_k c_k  \left(\bra{a} \opo^\dagger_k \opo_k \ket{a} + \bra{i} \opo^\dagger_k \opo_k \ket{i}- 2 \bra{a} \opo^\dagger_k \ket{a} \bra{i} \opo_k \ket{i} \right) \\
&\quad - \delta_{a i} \delta_{bj} \, \sum_k c_k  \left(\left|\bra{i} \opo_k \ket{j} \right|^2 +  \left|\bra{j} \opo_k \ket{i} \right|^2\right) \, .
\end{split}
\end{align}
Writing the approximate sizes of the matrix elements explicitly, this becomes
\begin{equation}
\bra{a i} H_{\rm TFD} \ket{b j}  = d_1 \, \delta_{ab} \delta_{ij} - d_2 \, \delta_{ai} \delta_{bj} \, ,
\end{equation}
where $d_1$ and $d_2$ are approximately equal and are of order $\opo(1)$. This implies that there is an eigenvalue at zero, and a bunch of eigenvalues coming from the second term separated from it by a distance $d_2$ roughly. The extra terms in the full Hamiltonian \eqref{eq:H_ele_basis} that we threw away are of order $\mathcal{O}(\sqrt{N})$. Thus they do not spread the eigenvalue spectrum by too much at large $K$, giving us a finite gap.

\begin{figure}[!htb]
\begin{center}
\includegraphics[scale = .8]{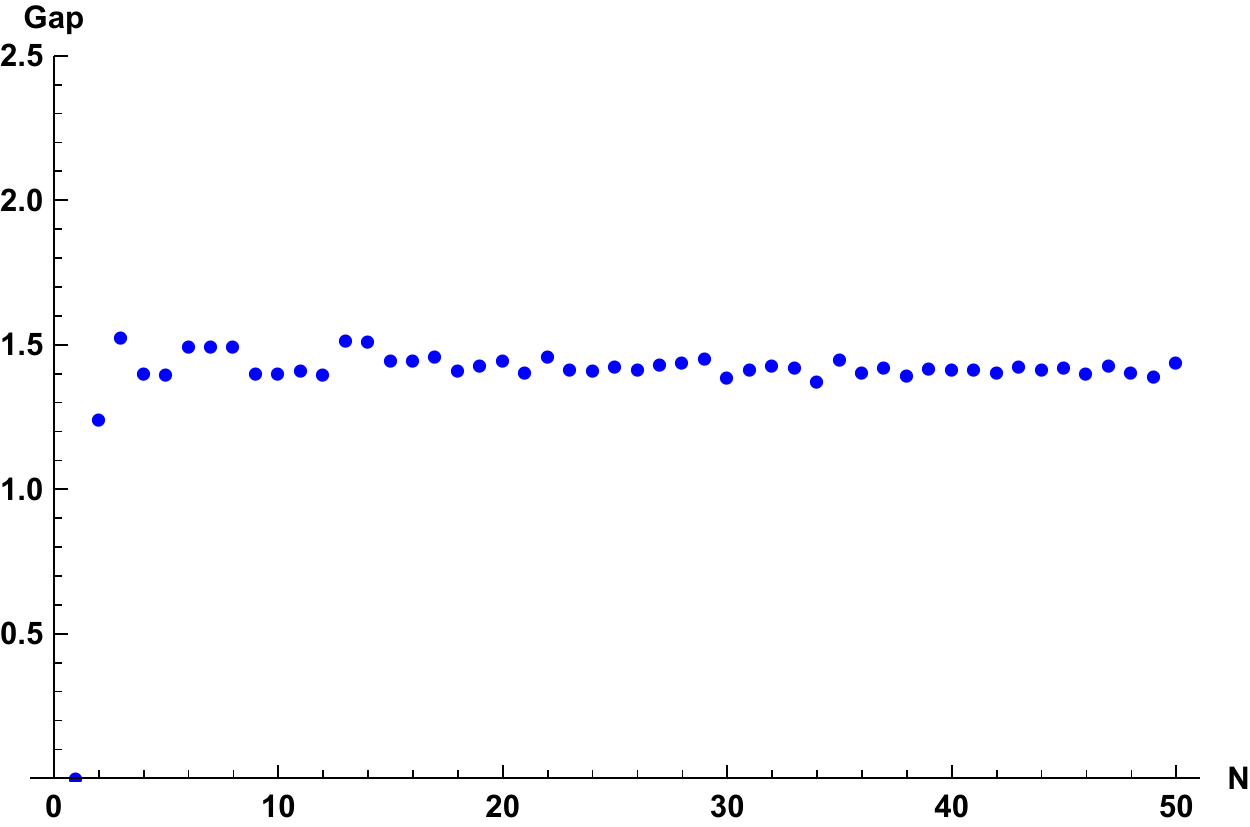}
\end{center}
\caption{Gap as a function of N for $\HTFD$ of dimension $\text{N}^2 \times \text{N}^2$ and K=50. }
\label{fig:GapvsNLargeK} 
\end{figure}

We can verify this estimate by studying the gap of the full TFD Hamiltonian \eqref{eq:genHelement} numerically. In Figure \ref{fig:GapvsNLargeK}, we plot this gap as a function of N, with the dimensions of the Hamiltonian being $\text{N}^2 \times \text{N}^2$.  

\begin{figure}[!htb]
$$
\begin{array}{cc}
  \includegraphics[angle=0,width=0.43\textwidth]{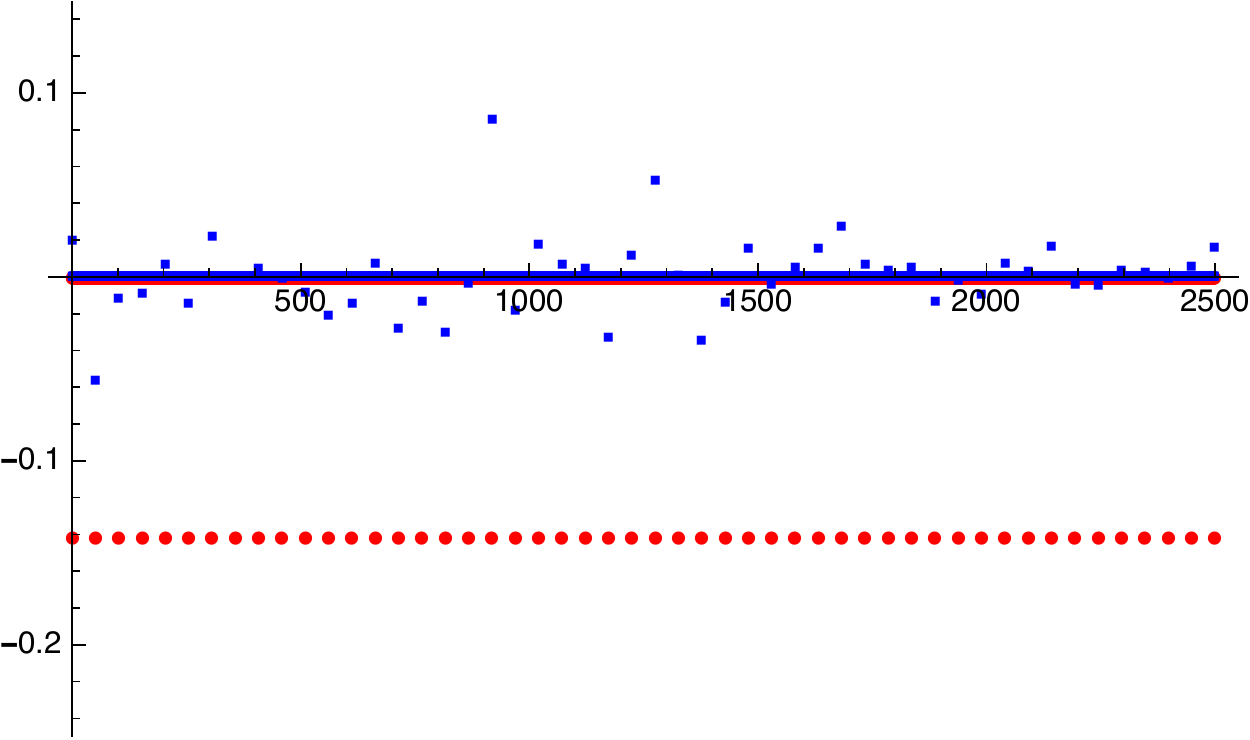} \qquad\qquad & \includegraphics[angle=0,width=0.43\textwidth]{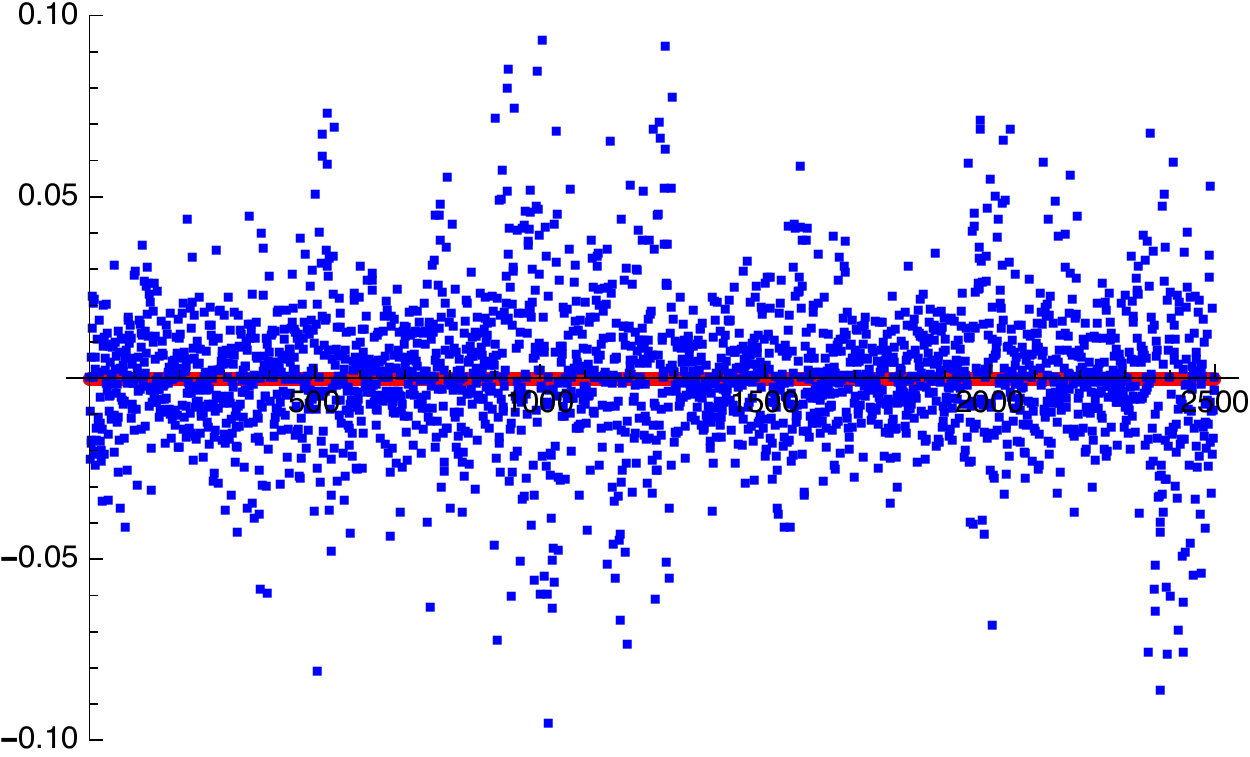}\\
  (a) \qquad\qquad & (b)
\end{array}
$$
\vspace{-0.7cm}
\caption{\small Components of the ground state (in red) and the first excited state (in blue) of the full Hamiltonian in (a) symmetric subspace and (b) the complement of the symmetric subspace. We have set K=N=50.}
\label{fig:EStates_windowLargeK}
\end{figure}

It is important to verify that the ground state of the TFD Hamiltonian in equation \eqref{eq:genHelement} is indeed the TFD state. Since we would like to construct an infinite temperature TFD state, analytically it is clear that the maximally entangled state in the energy window (with no phases) is the TFD state. The structure of the Hamiltonian in \eqref{eq:H_window1} implies that this is precisely the ground state. We can also verify this numerically. In Figure \ref{fig:EStates_windowLargeK} we plot the ground state and the first excited state of the full TFD Hamiltonian.

The ground state is plotted in red, the first excited state in blue. Part (a) shows the projection of these states in the symmetric subspace, and part (b) their projection in its complement. As we expect, the ground state (red) lies fully in the symmetric subspace, with a norm of 1.0000. The fact that it is constant in the symmetric subspace means it has an overlap of 1.0000 with the (infinite temperature) TFD state. The first excited state (blue) has little support in the symmetric subspace, with a norm of 0.1514. This supports the existence of a finite gap, as shown in Figure \ref{fig:GapvsNLargeK}.

\subsubsection{Small K}
\label{smallK}

We studied the eigenvalue spectrum of the TFD Hamiltonian in the case when $K$ is large (of the order of $N$). However, we would like to improve the situation by considering $K$ to be order $\mathcal{O}(N^0)$. There are two main motivations to do this. Firstly, it is expected that there are few light operators in holography. This comes from the sparseness of low-energy spectum of a holographic field theory. Thus it is more interesting to consider a TFD Hamiltonian with contributions from few operators. Secondly, it is more feasible to couple few operators in lab and such a scenario will then be easier to realize in a future experiment. 

\begin{figure}[!htb]
\begin{center}
\includegraphics[scale = 0.85]{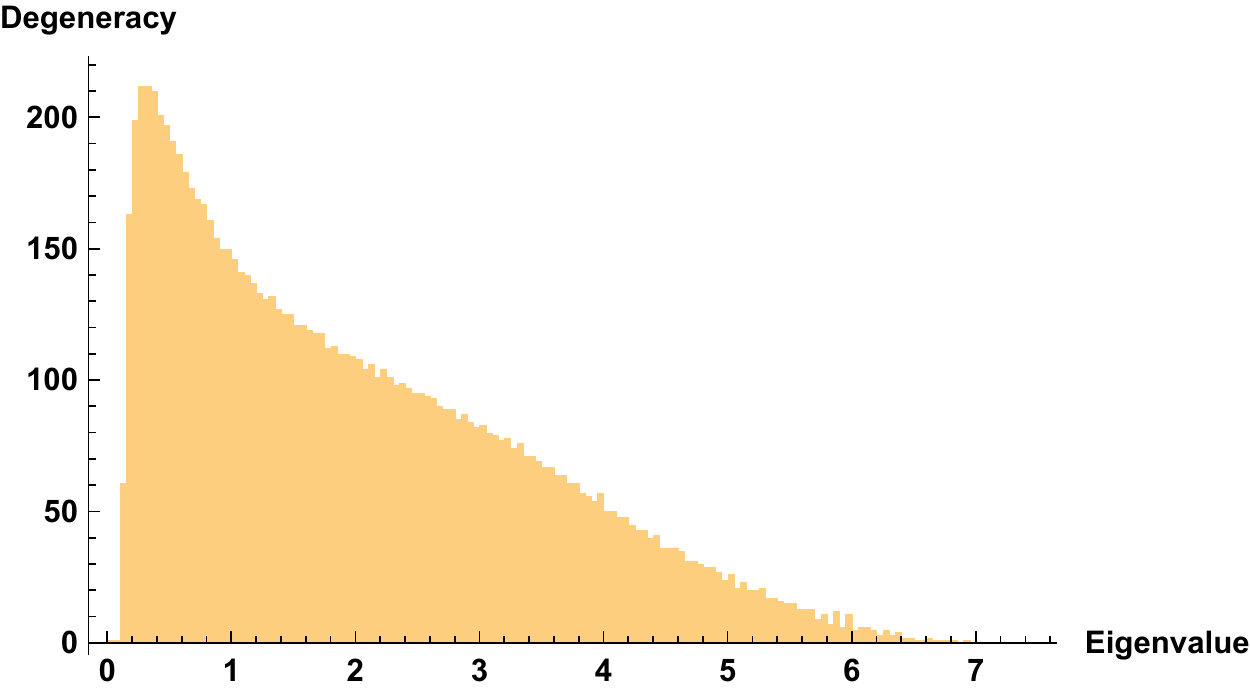}
\end{center}
\caption{Low-lying eigenvalue spectrum for $\HTFD$ of dimension $N^2 \times N^2$ with $N=100$, $K=2$. The gap is finite as seen in Figure \ref{fig:GapvsNK2}. }
\label{fig:LowEVDistK2}
\end{figure}

Now we recall the general matrix elements of the TFD Hamiltonian, equation \eqref{eq:genHelement}. We start by doing some numerical experiments to get some intuition for their behavior. In our toy model, we now choose $K=2$ and $N=100$. The energy eigenvalues of the original theories are again given by equation \eqref{eq:energy_func} and the operators by equation \eqref{eq:toy_ETH}.  With these choices, a typical eigenvalue distribution of the TFD Hamiltonian looks like Figure \ref{fig:LowEVDistK2}.

\begin{figure}[!htb]
$$
\begin{array}{cc}
  \includegraphics[angle=0,width=0.43\textwidth]{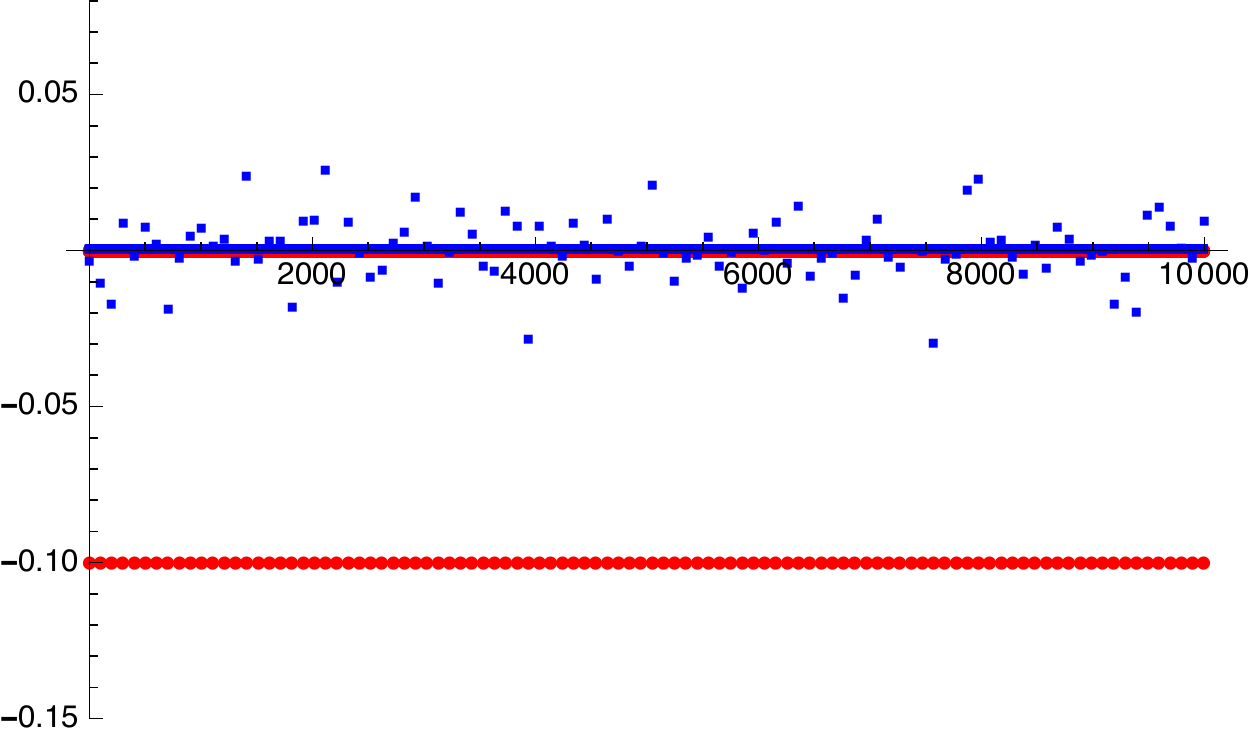} \qquad\qquad & \includegraphics[angle=0,width=0.43\textwidth]{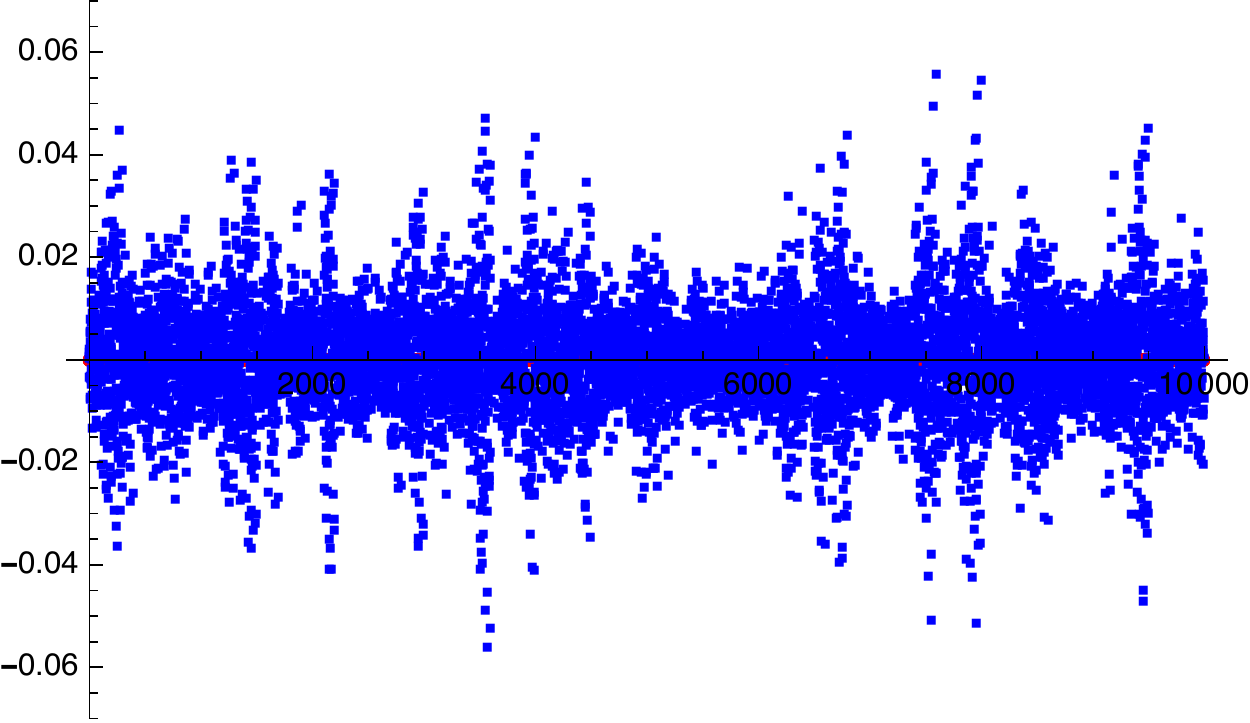}\\
  (a) \qquad\qquad & (b)
\end{array}
$$
\vspace{-0.7cm}
\caption{\small Components of the ground state (in red) and the first excited state (in blue) of the full Hamiltonian in (a) symmetric subspace and (b) the complement of the symmetric subspace. We have set N=100, K=2.}
\label{fig:EStatesK2}
\end{figure}

This is not a semi-circular distribution of eigenvalues, but the numerics do suggest the existence of a finite gap. The distribution is distorted from that in Figure \ref{fig:EVspectrumLargeK} because the off-diagonal matrix elements of the TFD Hamiltonian in equation \eqref{eq:genHelement} do not cancel each other. However, the ground state of the Hamiltonian is still the infinite temperature TFD state. We can test this in our toy model. In part (a) of Figure \ref{fig:EStatesK2}, we show the ground state (in red) and the first excited state (in blue) of the TFD Hamiltonian projected in the symmetric subspace. The norm of the projected ground state is 1.0000, with no relative phases. We also plot the first excited state, whose norm in the symmetric subspace is only 0.0997. 

\begin{figure}[!htb]
\begin{center}
\includegraphics[scale = .8]{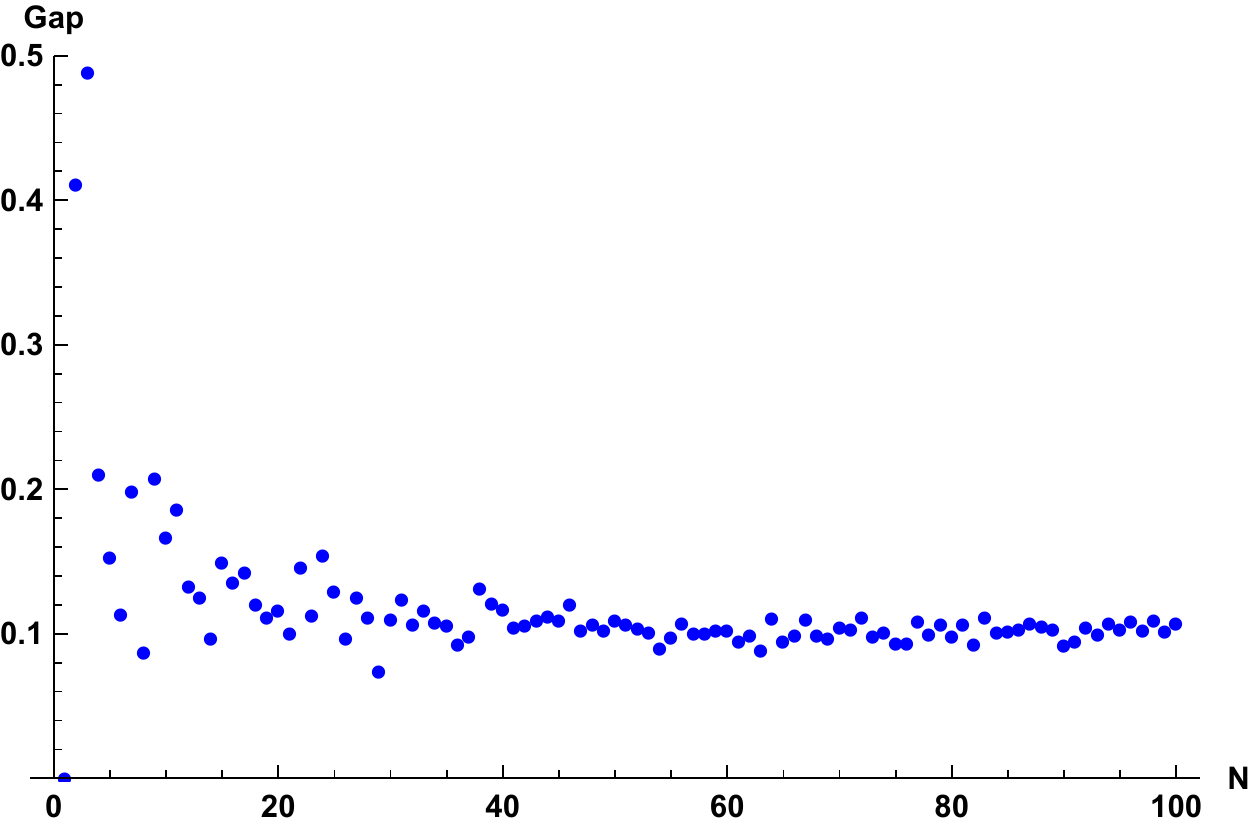}
\end{center}
\caption{Gap as a function of N for $\HTFD$ of dimension $\text{N}^2 \times \text{N}^2$ with $K=2$.}
\label{fig:GapvsNK2}
\end{figure}

  This further supports the finite gap one sees in the histogram \ref{fig:LowEVDistK2}. In fact, one can numerically study the behavior of gap as a function of $N$ and confirm that there is a finite gap of order $\mathcal{O}(N^0)$. We show this in Figure \ref{fig:GapvsNK2}.

It is tantalizing to imagine that there could be an analytic formula for the eigenvalue distribution shown in Figure \ref{fig:LowEVDistK2}. In fact, the distribution does look like some deformation of Marchenko-Pastur Distribution. But unfortunately, we were not able to complete the analytic study of the eigenvalue spectrum of the TFD Hamiltonian with contributions from few operators. We leave this to future work.

\subsection{Gap equation in full system}
\label{full_Hspace}

We now want to extend the reasoning above to the full spectrum of the theory. In particular,
 we want to take into account that the operators being used in the construction only connect states whose energy difference is not too big. We saw above that the TFD is the unique ground state within an energy window, and that the gap is not small. However, the TFD state has support over a wide range of energies, so there may be other states that look like the TFD within a small energy window but are orthogonal on the whole Hilbert space. Since the operators of our TFD Hamiltonian are somewhat local in energy space, these new states could be light. New light states would be a major obstacle to using our TFD Hamiltonian to cool to the ground state.
  
 To analyze this, in this section we calculate the gap for these states that look like the TFD in a narrow window of energies but have arbitrary dependence on the energy on longer scales.
Making use of the above results, we focus only within the diagonal subspace and use the standard ETH ansatz (\ref{ethans}).

We begin with the full finite-temperature Hamiltonian, in the form
\begin{align}
H = \sum_k c_k d^\dagger_k d_k \, , \quad  d_k = e^{-\gamma (H_L^0 + H_R^0)} \left( \opo_L - \opo_R^T \right)e^{\gamma (H_L^0 + H_R^0)} \, .
\end{align}
where $ \gamma = \beta/4 $. Focussing on the symmetric subspace, the eigenvalue equation becomes
\beq
H_{aa} \xi_a + \sum_b H_{ab} \xi_b = \lambda \psi_a \, . 
\eeq
Here we have expanded the state in the energy basis, and 
\begin{align}
\begin{split}
H_{aa} &=  \sum_k c_k \sum_b e^{\alpha (E_a - E_b)} \left(  \left| \opo^k_{ab} \right|^2 +  \left| \opo^k_{ba} \right|^2 \right) \, ,\\
H_{ab} &= -\sum_k c_k \left( \left|\opo_{ab}^k \right|^2 +  \left|\opo_{ba}^k \right|^2 \right) \, .
\end{split}
\end{align}
Note that this $H_{ab}$ term is nonzero for $a=b$, so it must be combined with $H_{aa}$ to obtain the full diagonal matrix element.

One can verify that the thermofield double state is an eigenstate of this Hamiltonian with zero energy. Due to the form of the Hamiltonian as a sum of terms $d^\dagger d$ with positive coefficients, there cannot be negative eigenvalues, so the TFD is the ground state. 

This can also be verified by a numerical calculation of the eigenvalues and eigenvectors of the Hamiltonian. For example, suppose the energy eigenvalues of the original Hamiltonians are given by
\begin{equation}
\label{eq:energy_func}
E(i) = \delta \left[1+ \right(W(\eta \, i)\left)^{1/p} \right] \, ,
\end{equation}
\noindent where $W(\cdot)$ satisfies $x = W(x e^x)$. The density of states is defined as
\begin{equation}
\rho(E) \equiv \frac{d \, i (E)}{dE} \, ,
\end{equation}
which turns out to be
\begin{equation}
\rho(E) = \frac{p \, E^{(2p-1)}  \, e^{\left( E/\delta \right)^p-1}}{\delta^{2p} \, \eta  } \, .
\end{equation}
This density function mimics the density of states of a QFT and hence the motivation to choose the energy to have the functional form in \eqref{eq:energy_func}. 

\begin{figure}[!h]
$$
\begin{array}{cc}
  \includegraphics[angle=0,width=0.43\textwidth]{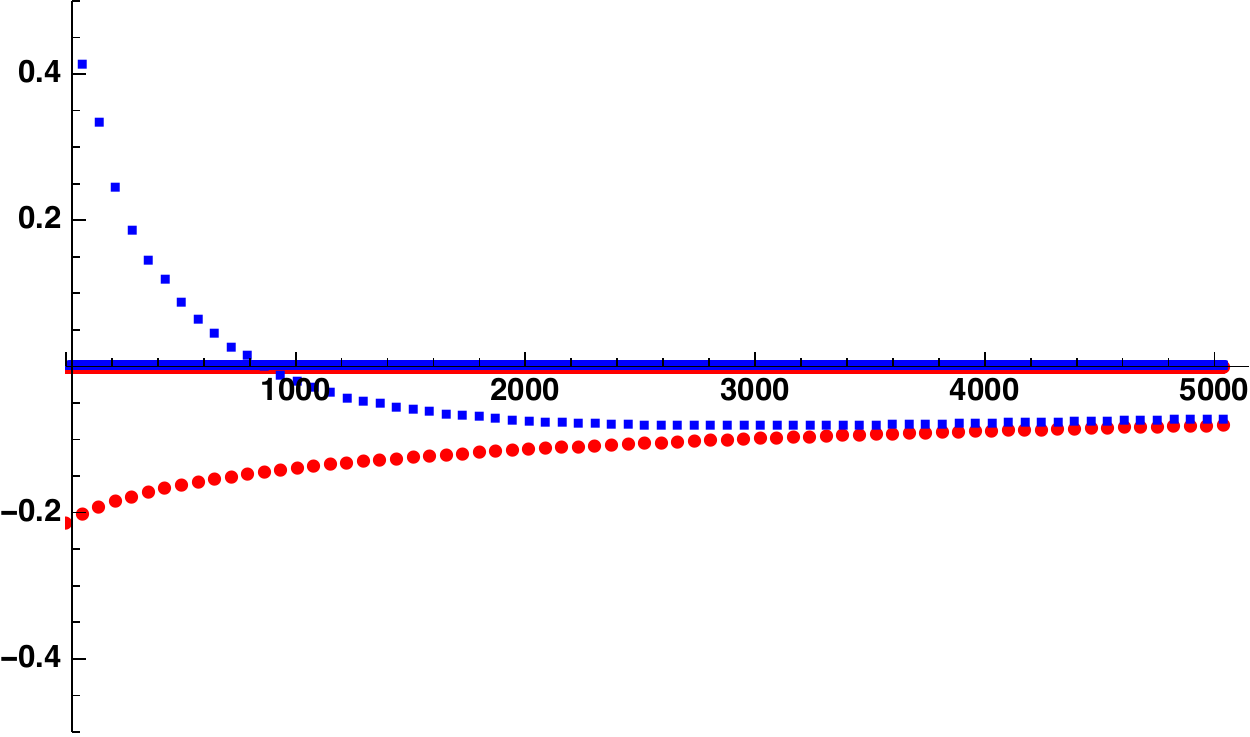} \qquad\qquad & \includegraphics[angle=0,width=0.43\textwidth]{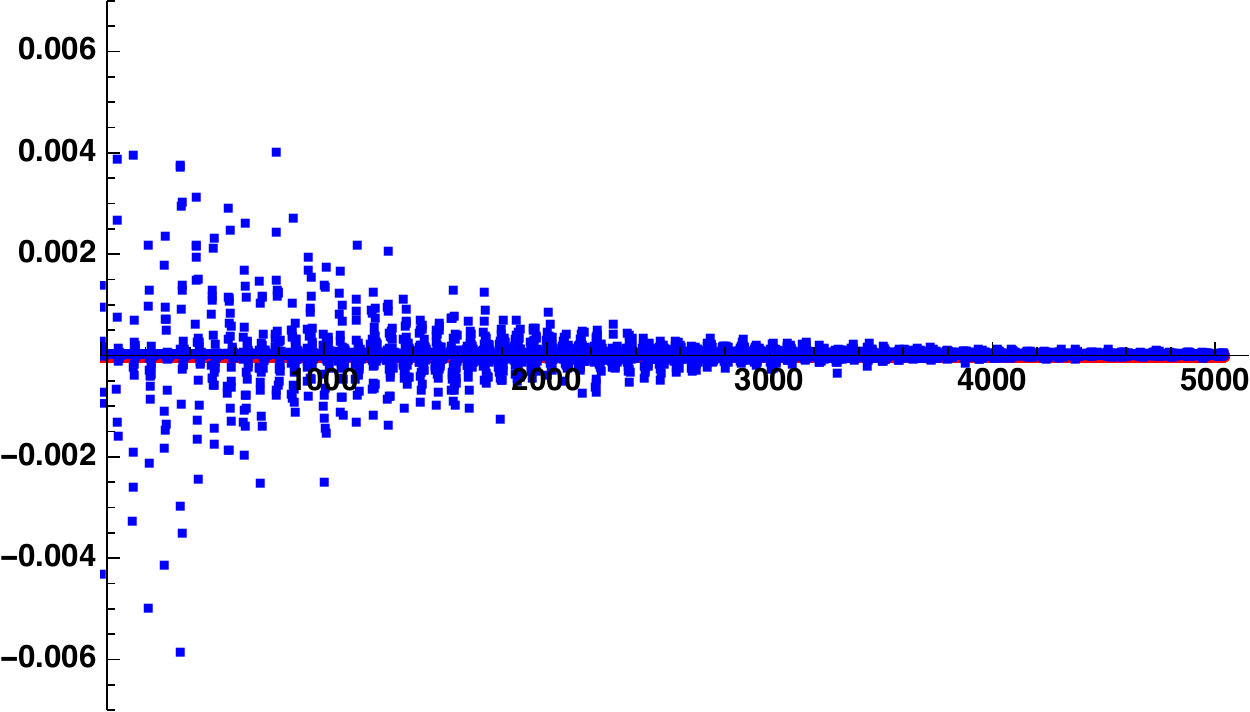}\\
  (a) \qquad\qquad & (b)
\end{array}
$$
\vspace{-0.7cm}
\caption{\small Components of the ground state (in red) and the first excited state (in blue) of the full Hamiltonian in (a) symmetric subspace and (b) the complement of the symmetric subspace in the full Hilbert space.}
\label{fig:FullHamStates}
\end{figure}

Using this model with the choices $\delta=1, \eta=1/3$ and $p=0.6$ we show in Figure \ref{fig:FullHamStates}  the numerical ground state and the first excited state in the symmetric subspace and in its complement in the full Hilbert space. The numerics support our expectation that the ground state is in the symmetric subspace, with its norm in the subspace being 1. Furthermore, even the first excited state is in the symmetric subspace, with a norm of 0.999657.
We would now like to estimate the gap between the ground state and the first excited state.

It is difficult to find the eigenvectors in general. However, it is natural to go to the continuum limit since the density of states is large in our ETH regime.

Define the state as
\be
\ket{\chi} = \sum_a  {\psi(E_a) \over \sqrt{\rho(E)}} \ket{a a} \, .
\ee

The eigenvalue equation becomes
\be
H_{aa} {\psi(E_a) \over \sqrt{\rho(E_a)}} + \sum_b H_{ab}  {\psi(E_b) \over \sqrt{\rho(E_b)}} = \lambda {\psi(E_a) \over \sqrt{\rho(E_a)}} \, .
\ee
Using the ETH ansatz, along with the ensemble average, 
\be
H_{ab} = - {1 \over \rho(\bar{E})} f(\bar{E}, \omega) \, ,
\ee
where $f$ is a positive function that depends strongly on the energy difference $\omega$ and weakly on the average energy $\bar{E}$. Converting from sums to integrals, the eigenvalue equation becomes
\be
\int d E' \rho(E') \left[ {1 \over \rho(\bar{E})}e^{\alpha (E - E')} f(\bar E, \omega) {\psi(E) \over \sqrt{\rho(E)}} - {1 \over \rho(\bar{E})}  f(\bar E, \omega) {\psi(E') \over \sqrt{\rho(E')}} \right] = \lambda {\psi(E) \over \sqrt{\rho(E)}}  \, .
\ee
Recall that $\bar{E} = {1 \over 2} (E + E')$ and $\omega = E'-E$. Define the density of states by
\be
\rho(E) = \Lambda e^{S(E)} \, ,
\ee
where $\Lambda$ is an arbitrary constant with dimensions of inverse energy. Then the eigenvalue equation becomes
\be
\int d E' f(\bar{E}, \omega) \left[ e^{S(E') - S(\bar{E}) - \alpha \omega} \psi(E) - e^{S(E')/2 + S(E)/2 - S(\bar{E})} \psi(E') \right] = \lambda \psi(E) \, .
\ee
To obtain the above equation, we have used the ETH ansatz, the statistics of our operators, and the continuum limit.

Now to make further progress, assume that the characteristic energy scale of our operators, $\Deltae$, is small compared to the total energy of the system, so the function $f(\bar{E}, \omega)$ is small unless the energy difference $\omega$ is small. Therefore, we can Taylor expand the entropy as a function of energy. In fact, we will see that the characteristic size of the energy difference $\omega$, which is set by the energy scale $\Deltae$ of our operators, is small compared to many other scales in the problem. Therefore we expand everything using $\omega$ as a small parameter (we will check later that this approximation is self-consistent)
to obtain
\begin{align}
\begin{split}
\psi(E) \int d \omega &f(\bar{E}, \omega) \left[ \omega (S'(E)/2 - \alpha) + {1 \over 2} w^2(S'(E)/2 - \alpha) ^2 + S''(E)w^2/4 \right]  \\
&- \int d \omega f(\bar{E}, \omega) \left(\omega \psi'(E) + {1 \over 2}\omega^2 \psi''(E)\right) = \lambda \psi(E) \, .
\end{split}
\end{align}
Define two functions encoding the first two moments of our operators,
\begin{align}
\begin{split}
\label{eq:f-g-defs}
g(E) &\equiv \int d \omega f(E+ \omega/2, \omega) \omega \, ,\\
h(E)^2 &\equiv  \int d \omega f(E+ \omega/2, \omega) \omega^2/2 \, .
\end{split}
\end{align}
With these definitions, the eigenvalue equation is simply
\be
- h^2 \partial_E^2 \psi - g \partial_E \psi + \left[{1 \over 2} (S'(E) - \beta) g + {1 \over 4} (S'(E) - \beta)^2 h^2 + {1 \over 2}S''(E) h^2 \right] \psi = \lambda \psi \, .
\ee
An important consistency check is that this operator should be Hermitian. Since we absorbed appropriate factors of the density of states into the definition of the wavefunction, the inner product is simply $ \int dE \,\psi_1^* \psi_2 $. Hermiticity then requires that our two functions obey the consistency condition
\be
g = \partial_E(h^2) \, .
\ee
This is not obvious from the definitions, but note that $f(E,\omega)$ is an even function of $\omega$, and by assumption has a mild dependence on its first argument. Therefore we can expand
\begin{eqnarray}
g(E) &\approx& \int d\omega \left[ f(E, \omega) \omega + \partial_E f(E, \omega) \omega^2/2 \right] =\int d\omega  \partial_E f(E, \omega) \omega^2/2 \, ,\\
\partial_E(h^2) &\approx& \int d\omega \partial_Ef(E, \omega) \omega^2/2 \, .
\end{eqnarray}
Therefore everything is consistent as long as the operators we are using have a mild dependence on the average energy, which is expected. In the following we will freely substitute $g = \partial_E (h^2)$.

In order to find the eigenvalues, it is convenient to redefine so that it is a standard Schrodinger equation. Define a new independent variable $y$ and rescale the wavefunction as follows:
\begin{align}
\begin{split}
\label{eq:new-vars}
dy = {dE \over h(E)}, \qquad  \psi =  {\bar{\psi} \over \sqrt h}  \, .
\end{split}
\end{align}
This leads to an equation for the rescaled wavefunction
\be
-{\partial^2 \over \partial y^2} \bar{\psi} + V \bar{\psi} = \lambda \bar{\psi} \, .
\ee
The potential $V$ takes the supersymmetric form
\be
V = W^2 + \partial_y W \, .
\ee
The superpotential is
\be
W = {1 \over 2} \left( \partial_E h + h S'(E)  - \beta h \right) \, .
\ee
It is also useful to note the potential in terms of the more intuitive variable $E$,
\be
V =  {1 \over 4} (\partial_E h)^2 + {1 \over 2} h \partial_E^2 h +  (S'(E) - \beta) h \partial_E h + {1 \over 4} (S'(E) - \beta)^2 h^2 + {1 \over 2}S''(E) h^2  \, .
\ee
In analyzing these equations, keep in mind that the independent variable $y$ of the Schrodinger equation is different from $E$.

Now we can calculate the spectrum. The ground state wavefunction has zero energy and  is given by $\bar{\psi} =  \exp(\int W dy)$. The integral can be done explicitly, giving
\be
\int W dy = \int W dE/h = {1 \over 2} \left( \log h  + S - \beta E \right) \, .
\ee
Therefore, explicitly
\be
\bar\psi = \sqrt{h} \exp\left[{1 \over2} \left(S - \beta E \right) \right] \, ,
\ee
which is the TFD state disguised by our redefinitions. 

It is crucial that this ground state is normalizable from the perspective of this Schrodinger equation. This is the case as long as we choose our operators such that at large energy, the function $h(E)$ grows slower than exponentially in the energy, a mild requirement. 

We will see that it is a good approximation to expand near $E_\beta$ to obtain
\be
\psi_{\rm TFD} \approx \# \exp\left[{1 \over 4} S''_\beta (E - E_\beta)^2 \right] \, .
\ee
Since $S''(E)$ is negative, the Gaussian has the correct sign.
Higher order terms in the exponential that we have neglected are  given by  $\partial^n_E S (\Delta E)^n$ with $n > 2$

As long as the entropy as a function of energy has a simple form such as a power law,
\be
S'' \sim {S \over E^2} \, ,
\ee
so that the characteristic spread of the wavefunction is given by
\be
{\Delta E \over E} \sim {1 \over \sqrt{-S''}  E_\beta} \sim {1 \over \sqrt S} \, .
\ee
The fluctuations around the mean energy are suppressed as $1/\sqrt{S}$ at large entropy. This behavior is familiar from statistical mechanics, and had to happen because the probability distribution for the energies in the TFD state is given by the standard canonical ensemble.

We can now check that the Gaussian approximation is good, continuing to assume that the entropy as a function of energy takes a simple power law form. The terms we have neglected are
\be
\partial^n_E S (\Delta E)^n \sim S \left( \Delta E \over E \right)^n \sim S^{1 - n/2} \, , 
\ee
where in the last equation we have used the Gaussian formula for $\Delta E$. Since we are working at large entropy $S$, we are justified in dropping terms with $n>2$.

In addition, we can now check whether our approximation of small energy difference is self-consistent. We require
\be\label{ineq}
\Deltae \ll  (\partial_E^2 S)^{-1/2} = \beta^{-1} \sqrt{C_V} \, ,
\ee
where $C_V$ is the heat capacity at the teperature $\beta^{-1}$.
This is easily satisfied for large systems because  the heat capacity is extensive in the size of the system.

\subsection{Solving for the gap.}

We can take advantage of the SUSY quantum mechanics structure in order to estimate the gap. Note that under our assumptions, the ground state of the original potential, $\bar\psi = \exp(\int W dy)$ is normalizable. The `partner potential'
\be
\tilde V = W^2 - \partial_y W \, ,
\ee
shares all of the energy eigenvalues with the original potential, except for the ground state. The wavefunction $\bar\psi = \exp(-\int W dy)$ is formally an eigenstate with eigenvalue $0$, but it is non-normalizable under our assumptions.

Therefore, the ground state energy of the partner potential $\tilde V$ is the same as the first excited state of the original potential. Since the ground state of the original potential has eigenvalue $0$, the gap is simply given by the ground state energy of the partner potential. Explicitly the partner potential is
\be
\tilde V =   {h^2\over 4} \left( \partial_E S - \beta \right)^2 - {1 \over 2} h^2 \partial^2_E S + {1 \over 4} \left(\partial_E h\right)^2- {1 \over 2} h \partial^2_E h  \, .
\ee
It is worthwhile considering how the different terms in this potential scale with the volume in a large system. Assuming we choose the number of operators to scale with the volume, $h^2 \sim V$. At high temperatures, $E \sim V$ and $S \sim V$ so $\partial^2_E S \sim V^{-1}$. Therefore, in the partner potential $\tilde V$, the first term scales linearly with the volume, while the second term is volume independent and the remaining terms depend inversely on volume. The same analysis holds if instead of large volume we consider large entropy or central charge.

At large volume, it is therefore sensible to ignore the last two terms and treat the second term as a perturbation of the first. Furthermore, we can expand around energy $E_\beta$ defined by
\be
S'(E_\beta) = \beta \, ,
\ee
to obtain
\be
\tilde V \approx {h^2 \over 4} S''^2_\beta (E-E_\beta)^2 - {1 \over 2} h^2 S''_\beta \, ,
\ee
where $S''$ denotes the second derivative of entropy with respect to energy evaluated at energy $E_\beta$  corresponding to temperature $\beta^{-1}$. Note that $S''$ is typically negative and is related to the heat capacity.

Our potential therefore becomes approximately quadratic with an overall shift. Calculating the ground state gives simply
\be
{\rm Gap} \approx h^2_\beta \left| S''_\beta \right| \, .
\ee
In simple situations, the $S''$ term is typically of order
\be
\left| S_\beta'' \right| \sim {S \over E^2} \sim {\beta \over E} \, .
\ee
To see how the gap behaves we need to know the behavior of $h(E)$.  
 Using the definition of $f$, we can write  $h$ in terms of matrix elements of the operators,
\begin{eqnarray}
h(E)^2 =  \sum_{k, b} c_k {1 \over 2} (E_b - E_a)^2 \left( \left|\opo_{ab}^k \right|^2 +  \left|\opo_{ba}^k \right|^2 \right) {\rho(\bar{E}) \over \rho(E_b)} \, .
\end{eqnarray}
Evaluating this at $E_\beta$ and expanding for small energy difference gives
\be
h^2_\beta \approx \sum_{k, b} c_k {1 \over 2} (E_b-E_\beta)^2  \left( \left|\opo_{ab}^k \right|^2 +  \left|\opo_{ba}^k \right|^2 \right) e^{-\beta (E_b-E_\beta)/2} \, ,
\ee
where the state $\ket{a}$ has energy $E_\beta$.
This quantity has a simple description in terms of the Euclidean path integral
\begin{align}
\begin{split}
h^2_\beta &\approx {1 \over 2} \sum_k c_k \bra{E_\beta}\left( \dot \opo^\dagger_k(0) \dot \opo_k(\beta/2) + \dot \opo^*_k(0) \dot \opo^T_k(\beta/2) \right)  \ket{E_\beta} \, , \\
 &\approx {1 \over 2}\sum_k c_k \langle \left( \dot \opo^\dagger_k(0) \dot \opo_k(\beta/2) + \dot \opo^*_k(0) \dot \opo^T_k(\beta/2) \right)  \rangle_\beta \, .
\end{split}
\end{align}
where dot denotes the derivative with respect to Euclidean time, and the operators are separated by $\beta/2$ in Euclidean time. In other words, this can be thought of as a correlator where the operators are in the two different CFT's. 

To be more explicit, we need to be more specific about the theory. We will now focus on quantum field theories such that the sum over $k$ represents a sum over different space positions as well as different species of operators $\opo_k$. In field theories without a mass gap, or with a gap small compared to the temperature scale, we expect
\be
\left< \dot \opo^\dagger_k(0) \dot \opo_k(\beta/2) + \dot \opo^*_k(0) \dot \opo^T_k(\beta/2)  \right>_\beta
 \sim T^{2 \Delta + 2} \, ,
 \ee
 where $\Delta$ is the scaling dimension of the operator $\opo$. 
 
 Further, it is natural that the number of operators scales with the volume. In general, we could choose the separation between operators to be any length scale, but to keep things simple we choose to insert one operator per thermal volume (in other words, our operators are separated by $\beta$). We also use factors of $\beta$ to make the Hamiltonian have the correct dimensions. Putting these assumptions together gives
 \be
 h^2(\beta^{-1}) = b ({\rm Vol}) T^{d+3} \, ,
 \ee
 where $b$ is a dimensionless constant that adjusts the overall normalization of the Hamiltonian.

Using in addition that in CFT's 
\be
E \sim c \left( \rm Vol \right) T^{d+1} \, , \quad S \sim c \left( \rm Vol \right) T^{d}\, , \quad S'' \sim {1 \over  c \left( \rm Vol \right) T^{d+2} } \, ,
\ee
where $c$ is the central charge,
we get
\be
{\rm Gap} \sim {b \over c} T \, .
\ee
 The central charge in the denominator is a bit annoying, but it is far better than the naive guess  $\exp(-c)$. If we want an order one gap, we should either have $c$ terms in the Hamiltonian, or simply a large coefficient $b$ proportional to $c$.

\section{Approximate TFD Hamiltonians}
\label{approxH}

In the above section, we worked with a Hamiltonian whose ground state is exactly the thermofield double state. However, it may be difficult to implement this Hamiltonian because operators of the form $\exp(-\gamma H) \opo \exp(\gamma H)$ are generally difficult to compute in a strongly coupled theory. In addition, if we want a simple bulk dual with two asymptotic regions, we would like the Hamiltonian at high energy to be dominated by the original Hamiltonians of the left and right systems. This motivates us to consider a simpler Hamiltonian and study its ground state, as we will do below.

\subsection{Simplest Hamiltonian} 

The simplest Hamiltonian we are aware of that produces something close to the TFD state is
\begin{equation}
H_S = H_L^0 + H_R^0  + \sum_k c_k (\opo_L^\dagger - \opo_R^*) (\opo_L - \opo_R^T) \, .
\end{equation}
We want to find the ground state and gap for this Hamiltonian. Using ETH, we can argue that the low-energy eigenstates lie in the symmetric subspace, like in the previous section. Random statistics will then help us calculate the gap. In the symmetric subspace, the eigenvalue equation for the Hamiltonian $H$ simplifies to
\begin{equation}
H_{aa} \psi_a + \sum_b H_{ab} \psi_b = \lambda \psi_a \, ,
\end{equation}
with the expressions
\begin{align}
\begin{split}
\label{eq:simpleHij}
H_{aa} &=  2 E_a + \sum_k c_k \sum_b  \left(  \left| \opo^k_{ab} \right|^2 +  \left| \opo^k_{ba} \right|^2 \right) \, ,\\
H_{ab} &= -\sum_k c_k \left( \left|\opo_{ab}^k \right|^2 +  \left|\opo_{ba}^k \right|^2 \right) \, .
\end{split}
\end{align}
Note that there is a competition here: the interaction terms would be minimized by the infinite-temperature thermofield double state, while the free Hamiltonians are minimized in the vacuum. 

Assuming that the interaction terms are strong enough, the ground state will occur in the high energy regime where ETH works. The analysis is similar to what we did in the previous section, but slightly different. We again define the wavefunction by
\begin{equation}
\ket{\xi} = \sum_a  {\psi(E_a) \over \sqrt{\rho(E)}} \ket{a a} \, .
\end{equation}
With this choice of the eigenstate, the eigenvalue equation becomes
\begin{equation}
\sum_{k, b}c_k \left( \left|\opo_{ab}^k \right|^2 +  \left|\opo_{ba}^k \right|^2 \right) \left(\psi(E_a) -e^{-S(E_b)/2 + S(E_a)/2} \psi(E_b) \right) = (\lambda - 2 E_a) \psi(E_a) \, .
\end{equation}
Now almost everything can be expanded for small energy difference, except that we do not wish to assume  
\begin{equation}
S(E_a) - S(E_b) \approx {E_a - E_b \over T(E_a)} \, ,
\end{equation}
is small. However, $S''$ is still small compared to the characteristic energy differences as in (\ref{ineq}). Notice that for this inequality to be satisfied simultaneously  with the requirement of locality at the temperature scale we must satisfy:
\be
1 \gg \frac{T^2}{\sigma_E^2} \gg \frac{1}{C_V}
\ee
This means that locality pushes us into the thermodynamic limit. Now expanding the second term gives
\begin{align}
\begin{split}
e^{-S(E_b)/2 + S(E_a)/2} \psi(E_b) &\approx e^{-S'(E_a) \omega/2} \bigg( \left[1 - {1 \over 4} S''(E_a)\omega^2 \right] \psi(E_a) \\
&\quad + \omega \psi ' (E_a) + {\omega^2 \over 2} \psi''(E_a) \bigg) \, ,
\end{split}
\end{align}
where as before $\omega \equiv E_b - E_a$. In the continuum limit, one can then write the eigenvalue equation as the differential equation
\begin{equation}
- h^2(E) \partial_E^2 \psi - g(E) \partial_E \psi + \left( k(E) + 2 E +  {1 \over 2} h^2(E) S''(E) \right) \psi = \lambda  \psi \, ,
\end{equation}
with the functions $h$ and $g$ are defined in equations \eqref{eq:f-g-defs} and 
\begin{equation}
k(E_a) \equiv \sum_{k, b}c_k \left( \left|\opo_{ab}^k \right|^2 +  \left|\opo_{ba}^k \right|^2 \right) \left( 1 - e^{-S'(E_a) \omega / 2}  \right) \, .
\end{equation}
We now redefine the wavefunction to use a new-variable $y$ as in equations \eqref{eq:new-vars}. This simplifies the eigenvalue equation and brings it into the standard Schr\"odinger form
\begin{equation}
\label{eq:y_scheqn}
-{\partial^2 \over \partial y^2} \bar{\psi} + V \bar{\psi} = \lambda \bar{\psi} \, ,
\end{equation}
with potential given by 
\begin{equation}
V =  k(E) +2E + {1\over 2} h^2 S''    +  {1 \over 4}h'^2 + {1 \over 2} h h'' ~.
\end{equation}
Note that the last two terms in the potential are proportional to inverse powers of the volume and can be dropped; the first two terms dominate at large volume.

To determine the gap, the key part is thus the behavior of the function $k(E)$. Using $S'(E) \equiv \beta(E)$, we can write $k(E)$ in terms of finite temperature 2-point functions of the operators as follows
\begin{equation}
k(E) = \sum_k c_k \left(\bra{E} \opo_k^\dagger(0) \opo_k(0) -\opo_k^\dagger(-\beta(E)/2) \opo_k(0)  \ket{E}  + \opo \leftrightarrow \opo^T \right) \, .
\end{equation}
Using ETH, we can show that is simply the difference between a 2-point correlator where both operators are in the same copy and a correlator where the operators are in different copies of the theory,
\be
k(E) = \sum_k c_k\left( \langle \opo_k^\dagger(0) \opo_k(0)\rangle_\beta - \langle \opo_k^\dagger(0) \opo_k(\beta/2) \rangle_\beta + \opo \leftrightarrow \opo^T \right) \, .
\ee
We can estimate these 2-point functions explicitly if we take the case of 2D CFT. Since we have regulated our operators by smearing them in space/time, the first term will be fixed in terms of UV properties of these operators. The potential then becomes
\be
k(E) = \sum_k c_k  \left({\sigma_E^{2 \Delta}} - T^{2 \Delta_k} + \dots \right) \, .
\ee
Here $\sigma_E$ is set by the UV regulator of our operator and the ellipsis denotes terms that are suppressed by higher powers of $T/\sigma_E$. 

\begin{figure}[!h]
$$
\begin{array}{cc}
  \includegraphics[angle=0,width=0.43\textwidth]{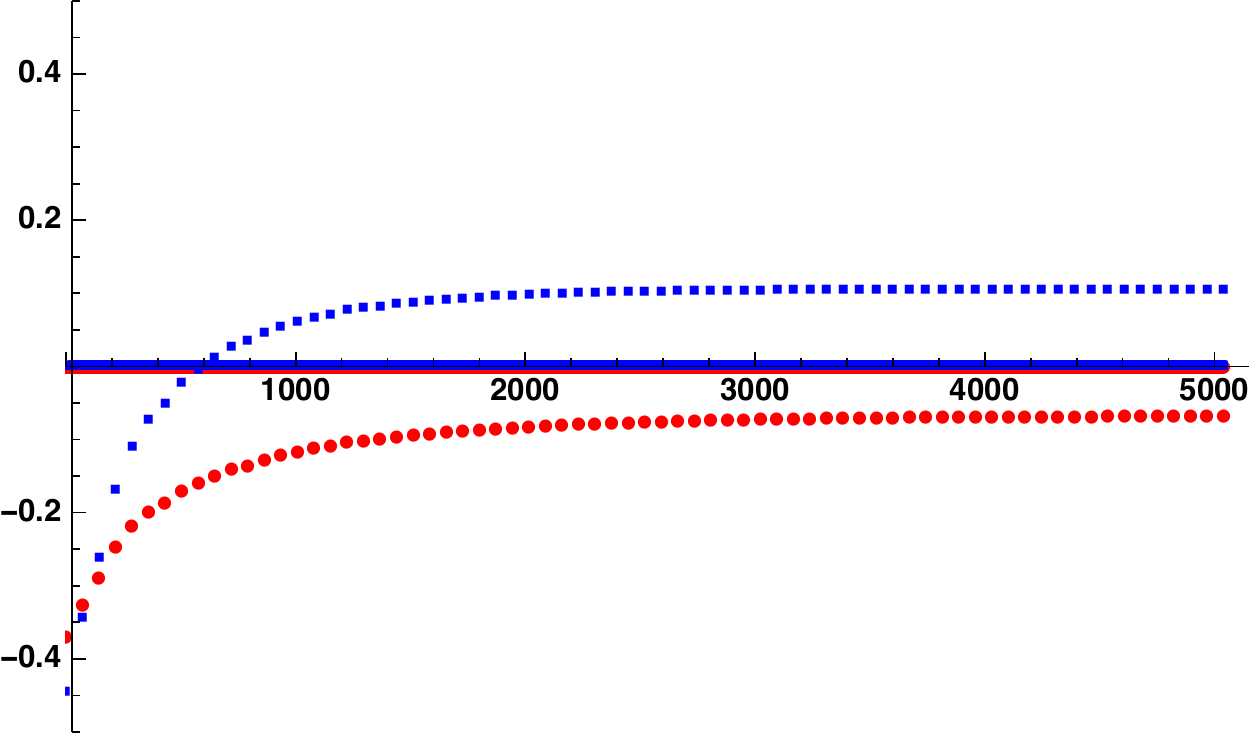} \qquad\qquad & \includegraphics[angle=0,width=0.43\textwidth]{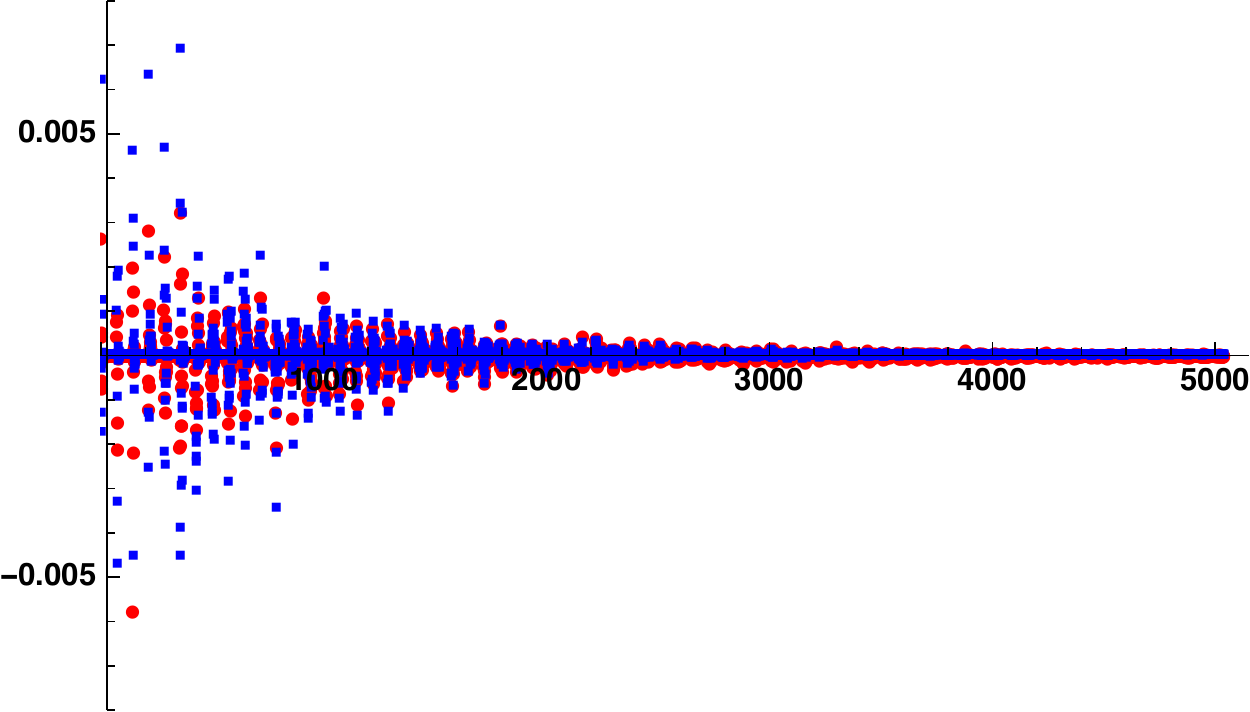}\\
  (a) \qquad\qquad & (b)
\end{array}
$$
\vspace{-0.7cm}
\caption{\small Components of the ground state (in red) and the first excited state (in blue) of the approximate Hamiltonian in (a) symmetric subspace and (b) the complement of the symmetric subspace in the full Hilbert space.}
\label{fig:ApproxHamStates}
\end{figure}

For simplicity, we assume that all of our operators have the same dimension $\Delta$; weakening this assumption will lead to minor modifications. We also assume a large $N$ limit for our theories. In this limit, the energy of a CFT is simply 
\begin{equation}
E \approx c V T^{d+1} \, ,
\end{equation}
where $d$ denotes the space dimension of the theory and $T= \beta(E)^{-1}$. The leading terms in the potential at large entropy will then be given by
\begin{equation}
\label{eq:simple_pot}
V(E) = \text{constant} + 2 c V T^{d+1} - \sum_k c_k T^{2 \Delta} + \dots \, .
\end{equation}
We now need the minimum of the potential $V(E)$, using the above estimate for $k(E)$. It is given by the condition
\begin{equation}
\label{eq:condition_ck}
{ (d+1) c \, V \over \Delta}  T_*^{d+1 - 2 \Delta} = \sum_k c_k \, .
\end{equation}
We now approximate the potential to be quadratic in energy near the minimum. The eigenvalue equation \eqref{eq:y_scheqn} implies that the gap is 
\begin{equation}
{\rm Gap} \approx \sqrt{\partial_y^2 \, V} \approx \sqrt{ h^2 \, \partial^2_E V }  \, ,
\end{equation}
where we have evaluated the second term at the minimum of the potential and used $\partial_E V =0$. It is easier to take $T$ derivative since the potential in \eqref{eq:simple_pot} is naturally defined in that variable. Doing this change of variables and assuming a large entropy, we get
\begin{equation}
\text{Gap} \sim h \, T_*^2 |S''| \sqrt{\partial_T^2 V|_{T_*}} \sim h T_*^2 |S''| \sqrt{c V T_*^{d-1}} \, ,
\end{equation}
where we have ignored order 1 constants like $d, \Delta$. Using now
\be
h = \sum_k c_k T_*^{2 \Delta + 2} \, .
\ee
Along with the condition \eqref{eq:condition_ck} yields
\be
{\rm Gap} \sim T_* \, .
\ee
  The ground state wavefunction is Gaussian. Going back to energy as our variable, we find

\be
\psi(E) \sim \exp\left[-\# |S''_*|  (E - E_*)^2 \right] \, .
\ee
In order to determine the order one number $\#$, we would need to know the functions $k(E)$ and $h(E)$ more precisely. Recall that the true thermofield double is given by the wavefunction
\be
\psi_{\rm TFD} \approx \# \exp\left[-{1 \over 4} \left|S''_\beta \right|(E - E_\beta)^2 \right] \, .
\ee
As long as our approximations hold, both the true and approximate TFD state live in the symmetric subspace, so they have the same entanglement structure. This can be easily verified to be the case numerically in the toy model we described near equation \eqref{eq:energy_func}. Figure \ref{fig:ApproxHamStates} shows the ground state and the first excited state of the approximate Hamiltonian in the symmetric subspace and in its complement in the full Hilbert space. The ground state is in the symmetric subspace to a very good approximation, as its norm of 0.99989 suggests. It has an overlap of 0.95445 with the exact TFD state. Further, even the first excited state is approximately in the symmetric subspace, with a norm of 0.99962.
However, the spread in energies may differ between the true and approximate TFD states by an order one factor determined by the unknown constant $\#$.  The entanglement entropy differs by an additive order one number, since the number of states differs by a multiplicative order one factor. Therefore, it is natural to conclude that the bulk dual of the approximate TFD state is also the eternal black hole. 

Finally, we would like to comment on how strong the `interaction term' in equation \eqref{eq:simpleHij} must be in order for the final temperature to be in the regime where ETH is valid. The interaction terms will be as large as the original Hamiltonian near the target temperature. For a CFT with large central charge $c$, we want the energy to be order $c$ to be in the ETH regime. If we use low dimension primary operators, the the interaction term $\left( \opo_L - \opo_R \right)^2$ is order one. Assuming that the number of different operators used is also order one, we require the coefficient $c_k$ of the interaction term to be of order the central charge,
\be
c_k \sim c \, .
\ee
This would make it difficult to calculate using perturbation theory in the bulk. However, our analysis above has not treated the interaction term perturbatively, so it should be valid in this regime.

\section{Discussion}

\subsection{Experimental Firewalls}
\label{sec-exp}

We have discussed in detail how to construct a TFD state as a ground state of a TFD Hamiltonian with a gap that is not very small. In this section, we will attempt to use a state such constructed to do some interesting gedanken experiments. In particular we will discuss what has been recently called a ``teleportation'' experiment. In the case of holographic field theories, our discussion will have an interpretation as probing firewalls in the eternal black hole geometry experimentally. 

We now briefly review the ``teleportation'' experiment as discussed in \cite{Gao:2016bin, Maldacena:2017axo} to set up the stage. Here one starts with two idential (holographic) field theories in the TFD state and perturbs the doubled-theory actively with a double-trace perturbation such that,
\begin{equation}
 S_{\text{tot}} \to S_{\text{tot}} + g \,  V \, ,
\end{equation}
with the constant $g$ chosen to have a specific sign before the experiment. The perturbation is of the form,
\begin{equation}
V \equiv {1 \over K}\, \sum\limits_{i=1}^K  \, \int d^{d-1} x \, \, O_L^i(0,\vec{x}) \, O_R^i(0,\vec{x})  \, .
\end{equation}
This looks very similar to interaction terms in the TFD hamiltonian that we have constructed. If we assume that the cooling procedure results in the TFD state at time $t=0$, such terms are already present in the total action. In the cases where the leading terms in the TFD Hamiltonian are the free Hamiltonians of the individual theories, these interaction terms can be thought of as a perturbation to the doubled-theory. If we decide to perform an experiment where $V$ is turned off a little while after $t=0$, physically separating the two theories so that they do not interact simulates this experiment as the interaction terms in the TFD Hamiltonian stop acting after the separation. The final result of the perturbation $V$ can then be measured by studying left-right correlators, for example. 

\cite{Maldacena:2017axo} studied such correlators using the dual eternal black hole geometry. The double-trace perturbation made up of many light operators then can be thought of as sending shockwaves into the bulk from the two boundaries. A signal from the left then interacts with the shockwave(s) and gives a non-zero commutator with a right operator. This was interpreted as traversing the wormhole. 

However, it is interesting to understand the role of $V$ in the experiment as well as in ``defining'' the TFD state. The maximally chaotic dynamics in the CFT does not care about the form of $V$ and thus the result of the experiment should be unaffected. But \cite{vanBreukelen:2017dul} showed that this is only the case when the signal is chosen appropriately. More precisely, if $V$ is replaced by $U_L^\dagger \, V \, U_L$, the signal that traverses the wormhole is $U_L^\dagger \, \phi_L \, U_L$ and not $\phi_L$. We can think of this as defining a ``dictionary'' between which left operators interacts with which right operators. Before the perturbation, such a dictionary is mere convention but it becomes physical once the perturbation opens the wormhole. We will see an example of this using the free fermion field theory in Appendix \ref{fermi_prop}.

\subsection{Quantum Computation and Machine Learning }

\label{qc_connection}
So far we have discussed the difficulty in constructing the TFD state, a specific state with maximal entanglement between two quantum systems, with particular focus on the gap in the TFD Hamiltonian. The problem of finding the minimal energy eigenvalue of the TFD Hamiltonian can be rephrased in the language of quantum computing. We will now discuss this interpretation in brief.

First, let us consider the `satisfiability problem', which is the first problem proven to be NP-complete. For orientation, first  consider the task to find obtain a specific value for Boolean variable $S \equiv x_1 \lor x_2$. Here, both $x_i$ are themselves Boolean variables, taking values $\{0,1\}$ and say we want to obtain $S$ with the value $1$. What are the possible combinations of values of $(x_1,x_2)$ that will ``satisfy" this task? The answer is well-known. The operation of disjunction (OR) we used above implies that $S$ is 1 whenever either or both of $x_1$ are 1. 

Now consider a more complicated problem. Let
\begin{equation}
S_i \equiv x_{\mu_1} \lor x_{\mu_2} \, \cdots \lor x_{\mu_k} 
\end{equation}
be a disjunction of $k$ Boolean variables $x_{\mu_i}$. Define the variable
\begin{equation}
S \equiv S_1 \land S_2 \land \cdots \land S_m \, .
\end{equation}
The problem to find one or more configurations of the base Boolean variables $\{ x_{\mu_i} \} $ that will satisfy $S=1$ is called a $k$-SAT problem in computer science. The decision problem of whether $S=1$ is possible can be shown to be in NP \cite{doi:10.1080/00107514.2018.1450720}. Moreover, it is known to be NP-complete for all $k \ge 3$. Such problems are critical elements for many other computational tasks.  

As described in  \cite{2016arXiv161207258P}, one may use a quantum computer to solve these statements via quantum annealing.  The strategy is to associate to each proposition a positive definite operator $H_{i}$ such that $H_{i}\ket{\psi}=0$ iff the state $\ket{\psi}$ obeys the logical statement defined by $S_{i}$.  The solvability of the SAT problem is then mapped to a question of whether $H = \sum_{i} H_{i}$ has a zero energy ground state.  

Quantum annealing-based algorithms have been used to implement $k$-SAT problems, in particular  the random SAT problem \cite{Biere:2009:HSV:1550723, 2003PhRvA..67b2314H}.  Here, the goal is to determine the probability that a given statement of a certain form is true.  The problem we have considered is related in an obvious way.   Rather than looking at classical boolean statements, we are considering the satisfiability of `quantum propositions' such as $d_{\mathcal{O}_{i}} \ketT=0$.  Our problem becomes the classical random satisfiability problem when we restrict to matrices having integral entries of the appropriate form. It would be interesting to explore further how the quantum version of random satisfiability relates to the classical one.

 Another avenue worth exploring is the relationship between our results and machine learning.  One could think of the structure of our $d$ operators as encoding a  mapping from the left to right system,
 \be
 \opo \to e^{-\beta H/2} \au \opo^\dagger \au^{-1} e^{-\beta H/2}~.
 \ee
 For the systems of interest this mapping is very complicated. 
 
 We can think of the operators $d_{\opo_k}$ as representing training data stated in an operator language. The system must then learn the full mapping.
 We have shown that a state annihilated by $d$ operators constructed from a set of operators $\opo_k$ will automatically be annihilated by $d$ operators form from any operator that can be generated from commutators of the $\opo_k$. We expect that generically a small number of operators $\opo_k$ will generate the entire algebra. Therefore, the system learns the correct mapping from a small amount of `training data.'
 
 Moreover, in this formulation of the machine learning problem we see that there is a deep connection between successful learning and the presence of firewalls: successful learning is encoded in a state with smooth horizon annihilated by all $d$ operators, where the mapping between left and right is encoded in the entanglement structure of the state.

In addition,  embedding machine learning problems into quantum mechanics may be of theoretical value since it allows one to phrase the problem of machine learning purely in the language of matrices.  Moreover, a `generic' learning problem should just reduce to the properties of large random matrices, about which much is known.

\subsection{Future Directions}
\label{sec-future}

We have provided reasonably strong evidence that given two copies of any quantum mechanical system obeying the Eigenvalue Thermalization Ansatz, a simple Hamiltonian exists whose ground state is the thermofield double state. This Hamiltonian generically has an energy gap of order the temperature.
A number of open questions and puzzles remain. 

\paragraph{Bulk Dual of TFD Hamiltonian.} We primarily thought of our TFD Hamiltonian as a means to prepare a particular state. However, suppose that we prepare the system in the ground state of the TFD Hamiltonian and then continue to evolve {\it with the TFD Hamiltonian.} In holographic examples, one would like to know the bulk dual. The system has time translation invariance. It has a coupling between the left and right CFT's, but in the case of interest the coupling is relevant, so in the UV the Hamiltonian factorizes.

The obvious guess is that the bulk dual is a static traversable wormhole with two AdS asymptotic regions. Indeed, this is what happens in the closely related construction of Maldacena and Qi  \cite{Maldacena:2018lmt} within the SYK model. However, there is a puzzle. Maldacena and Qi coupled a large number of fields between the two CFT's, with a coupling of order one. We have considered coupling a small number of fields, but with a coupling of order the central charge.

This is not a problem for our CFT analysis, since it is not perturbative in the coupling. But it does make the bulk dual mysterious\footnote{We thank Juan Maldacena for discussions on this point.}. For one thing, the strong coupling takes us out of the supergravity regime (see also \cite{Gao:2018yzk} where a similar issue arose) . In addition, existing constructions of traversable wormholes rely on Casimir energy to violate the Null Energy Condition. We expect that increasing the coupling between the two boundaries will only enhance the Casimir-type energy until the coupling reaches order one; stronger coupling would not be expected to allow for more negative Casimir energy with a small number of fields. 

For these reasons, we cannot, at this point, provide a gravitational description of our Hamiltonian.

\paragraph{Explicit Analysis in CFT.} We have shown explicitly in Section \ref{ising_model} how our construction works within the space of primaries of the Ising model. However, it would be very interesting to experiment with our simple Hamiltonian, and to do the full analysis including descendants. In some cases, the left-right interactions we add can be understood as irrelevant. This raises the interesting question: is there a simple UV complete theory that has the Ising model TFD as its ground state? A combination of the analysis presented in Section \ref{ising_model} and guesswork suggests that a plausible candidate for this is the theory
\be
\HTFD = H^0_L + H^0_R - a \int d  x \sigma_L(x) \sigma_R(x) - b \int dx \epsilon_L(x) \epsilon_R(x) \, , 
\ee
where $\sigma_{L, R}$ and $\epsilon_{L,R}$ are primary operators in the left and right CFTs, and $H^0$ is the Ising Hamiltonian on each side.

More generally, it is interesting to analyze our TFD Hamiltonian in strongly-coupled CFTs. In these theories, instead of using our ETH type arguments, one could use exact and statistical results for the OPE coefficients to diagnose whether the TFD is the ground state of a simple Hamiltonian. Our analysis in this paper suggests that it is, but this could be more rigorously shown or disproven using CFT results. 

In CFTs with a large number of primaries, a specific task is to check our claim that a small number of operators is sufficient to pick out the TFD state. We discussed the Commutator Property in Section \ref{sec_general_constr}. Based on the intuition gained from this property and some preliminary analysis, we expect a statement of the following kind to hold in general CFTs:

Given two primary operators in a CFT on the Riemann sphere, $O_1(z, \bar{z})$ and $O_2(z, \bar{z})$, with conformal dimensions $\Delta_i$ and spins $s_i$ respectively, if
\begin{equation}
d_1(z, \bar{z}) \, \ketT = d_2(z, \bar{z}) \ketT = 0 
\end{equation}
where \,  $d_i(z, \bar{z}) \equiv O_i^L(z, \bar{z}) -  e^{-\beta H/2} \, \Theta \, O_i^{R,\dagger}(z, \bar{z}) \, \Theta^{-1} \, e^{\beta H/2}  $; \, then
\begin{equation}
d_k(z, \bar{z}) \ketT = 0                                                                                                                   
\end{equation}
where  \, $d_k (z, \bar{z}) \equiv O_k^L(z, \bar{z}) - e^{-\beta H/2} \, \Theta \, O_k^{R,\dagger}(z, \bar{z}) \, \Theta^{-1} \, e^{\beta H/2}  $
and $O_k(z, \bar{z})$ is any operator that appears in the OPE of $O_1$ and $O_2$,
\begin{equation}
O_1(z, \bar{z}) O_2(w, \bar{w}) \sim \sum_k \frac{c_{ijk} \, O_k (z, \bar{z})}{(z-w)^{h_1+h_2-h_k} \, (\bar{z}-\bar{w})^{\bar{h}_1+\bar{h}_2-\bar{h}_k}} 
\end{equation}
with the condition that \, $ \Delta_k < \Delta_1+\Delta_2$ \, and \, $ s_k >  s_1+s_2-1  $. We haven't proved this statement yet, but we intend to return to this in future work.

\paragraph{Errors.} In order to move towards an experimental realization of our procedure, it is important to understand how robust our construction is against errors of various kinds. There could be different sources for errors. In the general definition of the TFD Hamiltonian \eqref{eq:HTFD_general}, if we fail to include enough number of operators $d_i$, we could get errors in the sense that the ground state will be highly degenerate and the actual state we prepare by cooling the system might be far from the TFD state. Another source of errors comes from the ambiguity in the form of the exact TFD Hamiltonian, discussed in Section \ref{sec_general_constr}. Different forms of the TFD Hamiltonian have different gaps. A problem could arise in this situation if the low energy spectrum is similar to  that of a  glass. And finally, there will be errors in the ground state due to practical constraints. These constraints could come from errors in the cooling technique one would use in experiments or from not letting the system cool for long enough time. One may hope to find a theoretical model to incorporate at least some of these errors and make quantitative statements, but we leave this for future work.

\section*{Acknowledgements}
We have benefited from many helpful discussions with colleagues over the course of this work, including J. de Boer, M. Cheng, J. Maldacena, X.-L. Qi,  S. Shenker, J. Stout, L. Susskind, B. Swingle, and E. Verlinde. DMH is supported in part by the ERC Starting Grant {\scriptsize{GENGEOHOL}}. BF is supported in part by the ERC Consolidator Grant {\scriptsize{QUANTIVIOL}}. SFL would like to acknowledge financial support from the Netherlands Organization for Scientific Research (NWO).

\appendix

\section{Teleportation with Fermions}
\label{fermi_prop}

Here we will consider the tensor product of two free fermion field theory and study the effect of a left-right double-trace-like interaction on the left-right correlator. This is an interesting example because the answer can be calculated analytically to all orders in $g$, the coupling constant of the double-trace term. Moreover, this case is relevant in the experimental implementation of our ideas in the Ising model. The total action of the doubled-theory looks like,
\begin{align}
 \begin{split} \label{fermiaction}
S &= S_{L} + S_{R} + S_{\text{int}}  \, , \\
S_{L,R} &= \int d^{d} x\,   \overline{\psi}_{L,R} \left(i\slashed{\partial} -m\right)\psi_{L,R} \, , \\ 
S_{\text{int}} &= \int d^{d} x \, A^{\mu} \left(\overline{\psi}_{L} \gamma_{\mu}\psi_{R} + \overline{\psi}_{R}\gamma_{\mu} \psi_{L} \right) \equiv g V \, .
 \end{split}
\end{align}
where the $S_{\text{int}}$ term can be thought of as being descended from (a Legendre transform) of the TFD Hamiltonian. The photon profile $A^\mu$ is a constant there and plays the role of a quench when we start the teleportation experiment. We take it to have the profile $A^{\mu} = \delta^{\mu 0} \alpha(t)$. We now want to calculate the Feynman propagator between the left and the right fields. For this, first we obtain equations of motion from the action (\ref{fermiaction}) :
\begin{equation}
\left(i \slashed{\partial} - m\right) \psi_{L,R}+\alpha(t) \gamma_{0} \psi_{R,L}=0 \, . 
\end{equation}
Now, switch to the basis $\psi_{\pm} = \psi_{L} \pm \psi_{R}$ and define $\psi_{\pm} = e^{\pm i \int_{-\infty}^{t} \alpha(t') dt'} \tilde{\psi}_{\pm}$.  The new field $\tilde{\psi}$ obeys the free equation of motion
\begin{equation}
\left(i \slashed{\partial} - m\right) \tilde{\psi}_{\pm}=0 \, .
\end{equation}
Thus, the system is integrable for any profile and  it is not difficult to compute tunneling amplitudes between the two CFTs exactly.   To do this first expand the transformed fields in terms of creation and annihilation operators:
\begin{equation}
\tilde{\psi}_{\pm} = \int \frac{d^{d-1}k}{(2\pi)^{d-1}}\frac{1}{2 w} \left(a_{\pm}^{s}(k) u^{s}e^{i k x} + b_{\pm}^{s\dagger}(k) v^{s} e^{-i k x} \right) \, , 
\end{equation}
where the ladder modes satisfy
\begin{align}
\begin{split}
\{a_{\pm}^{r}(p), &a^{s \dagger}_{\pm}(q) \} = \{b_{\pm}^{r}(p), b^{s \dagger}_{\pm}(q) \} = 4 w (2\pi)^{d-1} \delta^{(d-1)}(p-q) \delta^{rs} \, , \\ 
&\sum_{s} u^{s} \overline{u}^{s} = - \slashed{p} +m, \qquad \sum_{s} v^{s}\overline{v}^{s} = -\slashed{p}-m \, .
\end{split}
\end{align}
The extra factor of $2$ above comes about in going from the $L,R$ basis to the $\pm$ basis.  The thermofield double state in the new basis is
\begin{align}
\begin{split}
\ketT &= \frac{1}{\sqrt{Z}} e^{\int \frac{d^{d-1}k}{(2\pi)^{d-1}(2w)} e^{-\frac{\beta w}{2}}\left( a_{L}^{\dagger} a_{R}^{\dagger} + b_{L}^{\dagger} b_{R}^{\dagger}\right)} \ket{0,0} \, , \\ 
&\Rightarrow  \frac{1}{\sqrt{Z}} e^{-\frac{1}{2} \int \frac{d^{d-1}k}{(2\pi)^{d-1}(2w)} e^{-\frac{\beta w}{2}}\left( a_{+}^{\dagger} a_{-}^{\dagger} + b_{+}^{\dagger} b_{-}^{\dagger}\right)} \ket{0,0} \, .
\end{split}
\end{align}
All the time evolution will be absorbed into the operator insertions so we can work with the zero time $\ketT$.  The following formulas are useful and may be checked simply by expanding the exponent to at most quadratic order
\begin{align}
\begin{split}
\label{useful}
\braT a_{\pm}^{\dagger}(k) a^{\dagger}_{\pm}(q)\ketT &= 0 \, , \\ 
\braT a^{\dagger}_{\pm}(k) a_{\pm}(q) \ketT &= \rho_{f}(w) 4 w (2\pi)^{d-1} \delta^{d-1}(k-q) \, , \\ 
\braT a_{\pm}(k) a^{\dagger}_{\pm}(q) \ketT &= (1-\rho_{f}(w)) 4 w (2\pi)^{d-1} \delta^{d-1}(k-q) \ , \\ 
\braT a^{\dagger}_{-}(k) a^{\dagger}_{+}(q) \ketT &= e^{\frac{\beta w }{2}} \rho_{f}(w) 4 w (2\pi)^{d-1} \delta^{d-1}(k-q) \, , \\ 
\braT a_{+}(k) a_{-}(q)\ketT &= e^{\frac{\beta w}{2}}\rho_{f}(w) 4 w (2\pi)^{d-1} \delta^{d-1}(k-q)  \, ,
\end{split}
\end{align}
where $\rho_{f} = (1+e^{\beta w})^{-1}$ is the usual thermal fermion number density. Now let's consider the left-right Feynman propagator:
\begin{equation}
G_{LR}(x',x)\equiv i\, \bra{\text{TFD}} T \psi_{L}(x') \overline{\psi}_{R}(x) \ketT \, , 
\end{equation}
where $x \equiv (t, \vec{x})$.  The time ordering is just the usual time ordering.   Now, switch to the $\pm$ basis and compute using the formulas above:
\begin{align*}
\begin{split}
G_{LR}(x',x) &= \frac{i}{4}\, \langle T\left(\psi_{+}(x')\overline{\psi}_{+}(x)-\psi_{+}(x')\overline{\psi}_{-}(x) + \psi_{-}(x')\overline{\psi}_{+}(x) - \psi_{-}(x')\overline{\psi}_{-}(x)\right) \rangle \, , \\ 
&= \frac{i}{4}\, \Bigg{\langle} T\Bigg{(}e^{i (A(t')- A(t))} \tilde{\psi}_{+}(x')\overline{\tilde{\psi}}_{+}(x)-e^{ i (A(t') +A(t))} \tilde{\psi}_{+}(x')\overline{\tilde{\psi}}_{-}(x) \\ 
&+ e^{-i (A(t') + A(t))} \tilde{\psi}_{-}(x')\overline{\tilde{\psi}}_{+}(x) - e^{-i(A(t')-A(t))}\tilde{\psi}_{-}(x')\overline{\tilde{\psi}}_{-}(x)\Bigg{)} \Bigg{\rangle} \, .
\end{split}
\end{align*}
Using (\ref{useful}), we see that the first term is
\begin{align}
\begin{split}
\langle T&\left(\tilde{\psi}_{+}(x') \overline{\tilde{\psi}}_{+}(x)\right)\rangle \equiv \theta(t'-t)\langle \tilde{\psi}_{+}(x') \overline{\tilde{\psi}}_{+}(x)\rangle - \theta(t-t')\langle \overline{\tilde{\psi}}_{+}(x) \tilde{\psi}_{+}(x')\rangle  \\  
\langle \tilde{\psi}_{+}(x') \overline{\tilde{\psi}}_{+}(x)\rangle&= 2\int \frac{d^{d-1}k}{(2\pi)^{d-1}(2w)}\left( (-\slashed{k}+m)(1-\rho_{f})e^{i k(x'-x)} -(\slashed{k}+m) \rho_{f} e^{-i k(x'-x)} \right) \\ 
\langle \overline{\tilde{\psi}}_{+}(x) \tilde{\psi}_{+}(x')\rangle &= 2 \int \frac{d^{d-1}k}{(2\pi)^{d-1}(2w)}\left((-\slashed{k}+m) \rho_{f} e^{i k(x'-x)}-(\slashed{k}+m)(1-\rho_{f})e^{- i k(x'-x)}\right)  
\end{split}
\end{align}
Combining these into a single propagator using the $i \epsilon$ prescription we get
\begin{align}
\begin{split}
\label{g1}
\langle T&\left(\tilde{\psi}_{+}(x') \overline{\tilde{\psi}}_{+}(x)\right)\rangle \equiv 2 G_{0}(x,x') +  2 G_{ent}(x,x') \, , \\
&=  2\int \frac{d^{d}k}{(2\pi)^{d}}\Big{(} \frac{(-\slashed{k}+m) e^{i k (x'- x)}(1-\rho_{f}(w))}{k^{2} + m^{2} - i \epsilon} - \frac{(\slashed{k}+m) \rho_{f}(w) e^{-i k(x'-x)}}{k^{2}+m^{2} - i \epsilon}\Big{)} \, .
\end{split}
\end{align}
Here, $G_0(x,x')$ denotes the zero temperature Feynman propagator
\begin{equation}
G_{0} \equiv  \int \frac{d^{d}k}{(2\pi)^{d}} \frac{(-\slashed{k} + m) e^{i  k (x'-x)}}{k^{2} + m^{2} - i \epsilon} \, , 
\end{equation}
and $G_{\text{ent}}(x,x')$ denotes a piece induced by entanglement between the left and the right theory at finite temperature
\begin{equation}
 G_{ent} \equiv  - \int \frac{d^{d}k}{(2\pi)^{d}} \frac{\rho_{f}(w)\left( e^{ik(x'-x)}(-\slashed{k}+m)+e^{-ik(x'-x)}(\slashed{k}+m)\right)}{k^{2}+m^{2}-i \epsilon} \, .
\end{equation}
The formula for the $ - -$ propagator is identical, the only difference is the dressing by $e^{i A}$.   Next, it is easy to see that mixed correlators like $\Big{\langle}T \tilde{\psi}_{\pm}(x') \overline{\tilde{\psi}}_{\mp}(x)\Big{\rangle}$ are zero.  Plugging all this back into the expression (\ref{g1}) we find
\be
G_{LR}(x,x') = - \sin(\Delta A) \left(G_{0} +  G_{\text{ent}}\right) \, ,
\ee
where $\Delta A = A(t') - A(t) = \int_{t}^{t'} \alpha(s)ds$.  Again, the first piece is trivial and comes about simply due to the direct interaction.  The second is due to the entanglement.  

For entangled free theories, modified left-right correlators are enough to diagnose how the signal propagates from one theory to another after the perturbation. But for chaotic theories, a better diagnostic is the commutator,
\be
C = \bra{\Omega} e^{-i g V} [\phi_{R}(t_{R}),\phi_{L}(t_{L})] e^{i g V} \ket{\Omega} \, , 
\ee
where $\phi_{L,R}$ represents the operator we are trying to send through the wormhole, $\Omega$ is the vacuum state that we have created, and $t_{R}$, $t_{L}$ represents the time when we create/measure the operator. For our toy model, it is straightforward to calculate this using the method outlined above. It would be interesting to analyze what happens when $\ket{\Omega}$ is not exactly the TFD state, but we leave this for the future.

\bibliographystyle{jhep}
\bibliography{references}

\end{document}